\documentclass{article}
\usepackage{PRIMEarxiv}
\usepackage[utf8]{inputenc} 
\usepackage[T1]{fontenc}    
\usepackage{hyperref}       
\usepackage{url}            
\usepackage{booktabs}       
\usepackage{amsfonts}       
\usepackage{nicefrac}       
\usepackage{microtype}      
\usepackage{lipsum}
\usepackage{fancyhdr}       
\usepackage{graphicx}       
\graphicspath{{figs/}}      
\usepackage{float}
\usepackage{tcolorbox}
\usepackage{enumitem}

\usepackage{multirow}
\usepackage{pdflscape}
\usepackage{caption}
\usepackage{adjustbox} 
\usepackage[toc,page]{appendix}

\usepackage[style=apa, backend=biber]{biblatex}
\title{
    Financial Bond Similarity Search Using Representation Learning\\
}

\author{
    Amin Haeri, Mahdi Ghelichi\\
    Model Development, Risk Management\\
    TD Bank, Toronto, Canada\\
    \texttt{\{amin.haeri, mahdi.ghelichi\}@td.com}\\
    \AND
    Nishant Agrawal, David Li, Catalina Gomez Sanchez\\
    Model Development, Risk Management\\
    TD Bank, Toronto, Canada\\
    \texttt{\{nishant.agrawal, zhao.li, catalina.gomezsanchez\}@tdsecurities.com}\\
}

\begin{document}
\maketitle

\begin{abstract}
Finding similar bonds remains challenging in fixed-income analytics, as numerical financial attributes often overshadow categorical non-financial ones such as issuer sector and domicile. This paper shows that these categorical attributes dominate the predictability of spread curves and proposes embedding models to capture their semantic similarities, outperforming one-hot and many other baselines. Evaluated via sparse-issuer augmentation, the approach improves risk modeling and curve construction.
\end{abstract}


\section{Introduction} \label{sec:intro}
In fixed-income markets, bonds are debt instruments through which governments, corporations, and other entities obtain capital from investors. They specify contractual terms such as coupon rate, maturity, and redemption method, which define how interest and principal are repaid. The bond universe spans sovereign, corporate, municipal, and structured issues, each carrying distinct risk and return patterns shaped by credit quality, issuer profile, and market dynamics. Given this diversity, comparing bonds is essential for portfolio construction and risk management.

Similarity search provides a data-driven framework to identify bonds with comparable characteristics across large datasets. Building on decades of research in distance metric learning \parencite{metriclearning} and information retrieval \parencite{ir}, similarity-based approaches have proven particularly valuable in financial contexts where direct price discovery is limited due to market microstructure constraints. It assesses closeness not only in financial features such as yield, duration, and credit risk but also in contextual dimensions like issuer industry, credit rating, and descriptive information. For instance, an algorithmic comparison should recognize that a bond issued by the Province of Manitoba is more closely related to other Canadian provincial issuers than to a corporate bond issued by Amazon. Such methods improve the relevance of peer selection and enhance the reliability of bond curve construction when direct market data are limited.

In practice, non-financial categorical attributes such as issuer industry, domicile, and market of issuance often explain more of the cross-sectional structure in spreads than marginal differences in numerical terms such as coupon or residual maturity. For example, a 7-year bond issued by the Province of Manitoba tends to behave more like other Canadian provincial issues of similar rating than a corporate bond issued by Amazon with nearly identical maturity and coupon, despite the latter being numerically closer on standard term-structure dimensions. Traditional similarity engines usually start from numerical financial variables and treat categorical descriptors as afterthoughts or filters, which can underweight the role of issuer-level and sectoral context in shaping spreads. The central claim of this paper is that these categorical non-financial attributes should instead be treated as the primary drivers of bond similarity. Recent advances in representation learning \parencite{replearning, bert} demonstrate that learned embeddings of categorical features can capture semantic relationships more effectively than explicit one-hot encodings; we extend this finding to the fixed-income domain. Handling categorical information correctly is therefore more important than adding further numerical detail for similarity assessment in bond markets.

Representation learning is a machine learning paradigm focused on automatically discovering useful features or representations of data, rather than relying on handcrafted variables. The goal is to transform raw inputs such as text and structured financial attributes into compact, informative vectors that capture semantic, structural, and statistical relationships. These learned representations enable downstream tasks like classification, clustering, retrieval, or prediction to be performed more effectively and with less domain-specific engineering. Advances in deep learning have made representation learning especially powerful in financial contexts \parencite{dl, dlfinance}, with models such as autoencoders, word embeddings, and transformer-based architectures that capture rich contextual patterns. For high-cardinality categorical variables (such as bond issuers or industry classifications), learned embeddings overcome the curse of dimensionality inherent in one-hot encoding, producing continuous latent representations where semantic proximity is reflected by vector distance.

Conventional approaches to bond similarity search rely heavily on handcrafted features, rule-based filters, and simple distance metrics applied to a limited set of attributes, such as coupon, maturity, or credit rating. While straightforward, these methods often fail to capture the full complexity of bond structures and issuer-specific nuances, especially when unstructured information from prospectuses or covenant terms is involved. More fundamentally, for high-cardinality categorical variables (e.g., issuer identity, industry classification), one-hot encoding treats all categories as orthogonal and prevents the model from learning meaningful similarities between categories. These conventional approaches can be brittle when handling missing data, issuer sparsity, or evolving market conditions, leading to inaccurate or incomplete similarity assessments. They lack the flexibility to adapt to new issuers or market regimes, restricting their utility in dynamic fixed-income markets.

In this paper, rather than positioning our work as another supervised or unsupervised metric-learning approach, we focus on the distinction between financial and non-financial attributes and on how to represent high-cardinality categorical information. The proposed framework uses embedding models to map these categorical attributes into a latent space where cosine similarity reflects economically meaningful relationships, and evaluates their usefulness via CDS spread-curve estimation from sparsified issuer catalogs.

\begin{figure}[t]
    \centering
    \includegraphics[width=0.33\textwidth]{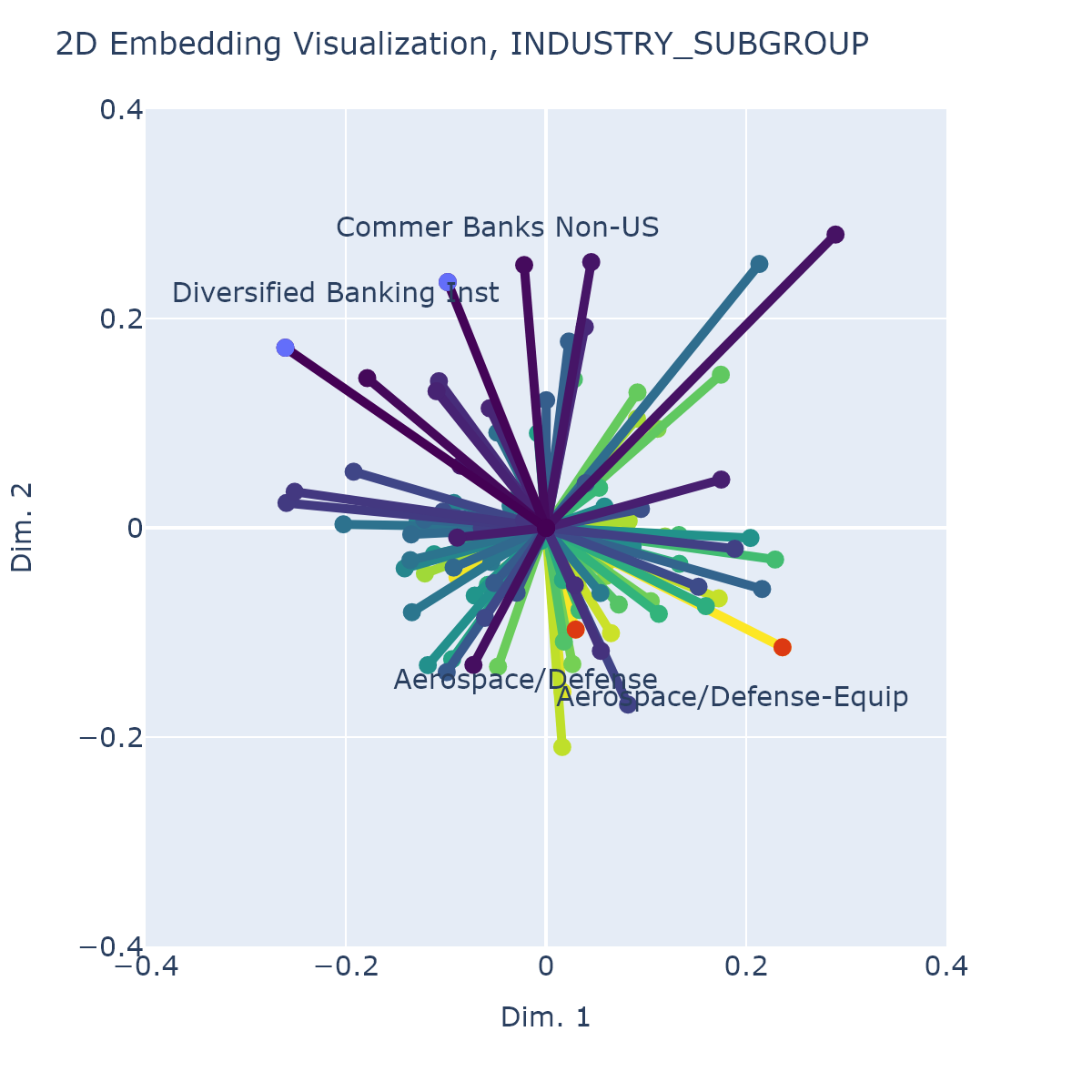}
    \includegraphics[width=0.33\textwidth]{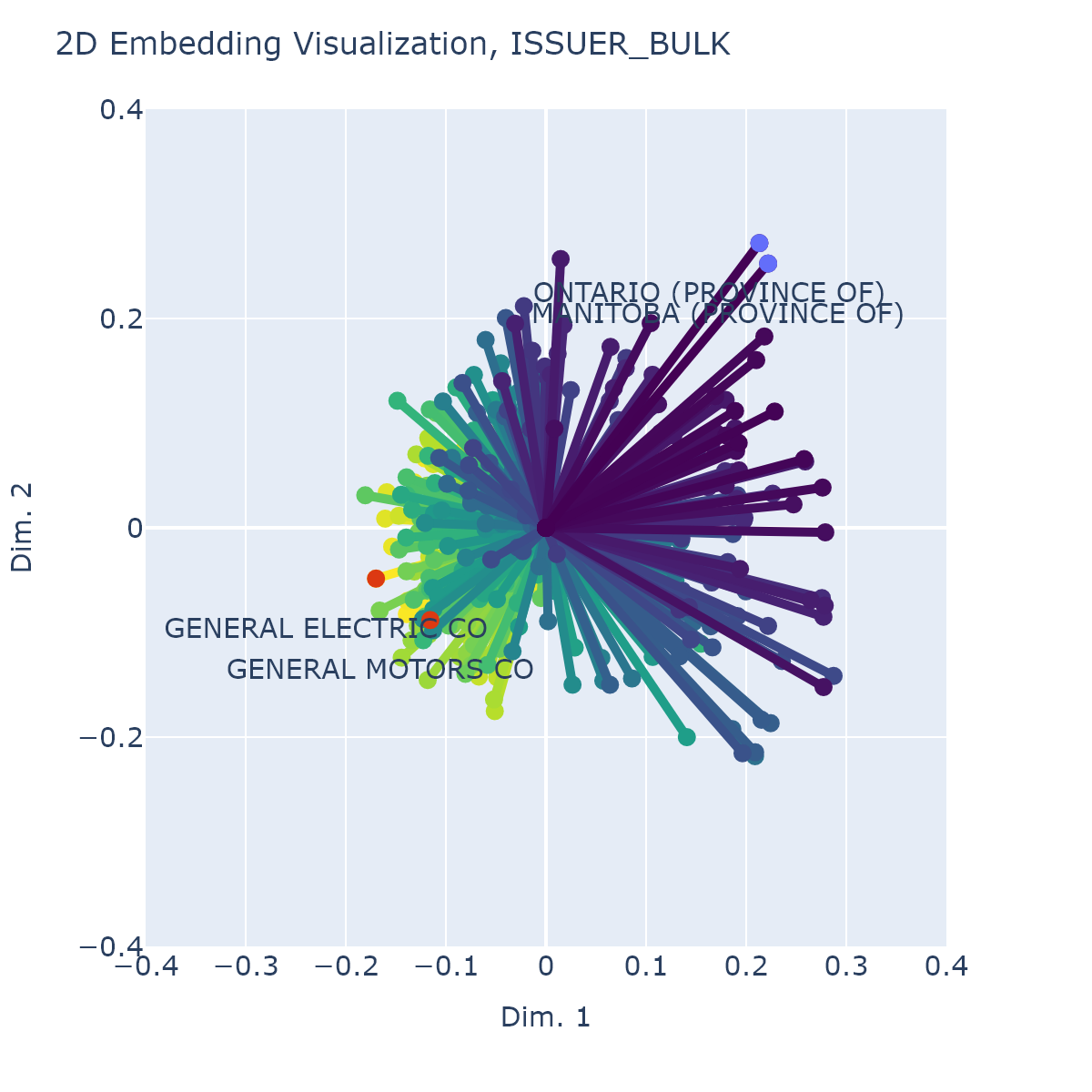}
    \includegraphics[width=0.33\textwidth]{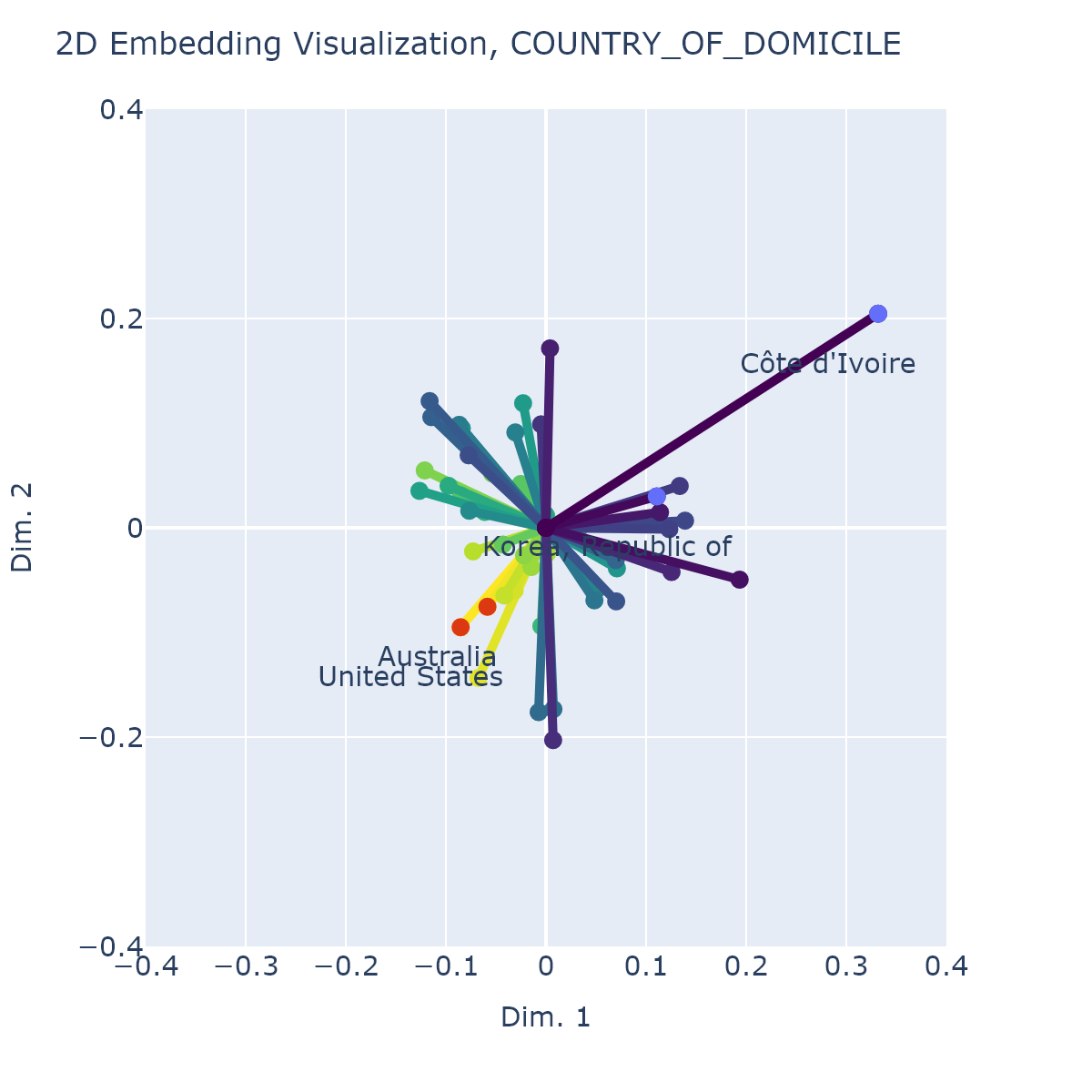}
    \caption{Two-dimensional embedding projections colored by their similarities in the high-dimensional space. From left to right: Industry Subgroup, Issuer Bulk, and Country of Domicile. Each point represents an entity in the learned embedding space, projected onto the first two dimensions, with radial lines indicating displacement from the origin. The visualizations illustrate how semantically related entities cluster and separate according to industry, issuer identity, and geographic domicile, highlighting the embedding's ability to capture structured relationships across different categorical views.}
    \label{fig:2demb}
\end{figure}

\section{Prior Work} \label{sec:pwork}

%
The majority of prior work on bond similarity and term-structure modeling has focused on numerical financial variables (i.e., yields, spreads, durations, and volatilities) combined with distance metrics or predictive models, while treating issuer-level and sectoral descriptors as coarse controls or exclusions. Unsupervised clustering and supervised metric learning have both been explored in this numerical regime, but the relative importance of non-financial categorical attributes and the question of how to represent them remain largely unaddressed.

Early research predominantly focused on unsupervised clustering methods such as K-means, hierarchical clustering, self-organizing maps, and Gaussian mixture models. These methods group bonds based on attributes like coupon rate, yield-to-maturity, duration, and credit rating.
They typically operate on standardized numerical features; industry or rating dummies, if used at all, are encoded via sparse one-hot vectors that do not express similarity between categories. While useful for broad classification tasks, such as distinguishing between investment-grade and high-yield bonds, unsupervised methods suffer from several limitations. They rely on predefined distance metrics (e.g., Euclidean or Minkowski), which may not align with the intrinsic geometry of financial data.

More recent work has shifted toward supervised similarity learning, where similarity is defined relative to a prediction target (e.g., yield or spread). In this paradigm, machine learning models are trained to predict bond characteristics, and proximity measures are extracted from the resulting model structure. Random forests, in particular, have been shown to be effective distance-metric learners, with theoretical justification grounded in the adaptive nearest-neighbors interpretation \parencite{rf, Scornet2015}. Proximity between two securities is defined by the proportion of trees in which they co-occur in the same terminal node. Variants such as out-of-bag (OOB) proximity \parencite{Breiman2004CONSISTENCYFA} and geometry-and-accuracy-preserving (GAP) proximities \parencite{JMLR:v17:14-168} have been proposed to improve robustness and interpretability. In financial applications, Random Forest-based similarities have proven scalable to high-dimensional datasets and naturally handle mixed categorical and numerical features \parencite{Desai}.

Building on these advances, quantum cognition machine learning (QCML) has recently been introduced as a novel paradigm for supervised distance metric learning \parencite{qcml}. QCML leverages the mathematical formalism of quantum theory and foundational ideas from quantum cognition \parencite{Busemeyer_Bruza_2012} by representing data points as quantum states in a Hilbert space, with features and targets encoded as quantum observables. Similarity is defined via quantum fidelity, a measure of overlap between two quantum states derived from quantum information theory, specifically the absolute value of the inner product between two quantum states. A key theoretical advantage of QCML is its logarithmic scaling of parameters with respect to feature dimensionality (the parameter count scales as $O(N^2)$, where $N$ is the Hilbert space dimension), in contrast to random forests, where tree depth typically scales linearly with the number of features, resulting in exponential growth in parameters and leaves. This logarithmic economy of representation endows QCML with superior generalization in high-dimensional, sparse data regimes. Empirical results demonstrate that QCML outperforms random forests in predicting yields and deriving similarity metrics for high-yield bonds, while performing comparably in investment-grade markets. This suggests that QCML may provide a robust alternative to tree-based methods for illiquid and noisy financial datasets, particularly when the bond universe exhibits high feature dimensionality (due to one-hot encoding of categorical variables) and a high concentration of outliers (as in high-yield markets).

While supervised similarity metrics based on tree ensembles and quantum models excel in capturing complex nonlinear relationships through prediction tasks, they remain bound by the representations inherent in their training data. An orthogonal but complementary line of inquiry is representation learning for categorical variables themselves. Transformer-based embeddings have demonstrated remarkable ability to capture semantic relationships in high-cardinality categorical spaces \parencite{transformers, bert}. However, their application to financial similarity, particularly in the context of bond markets where categorical non-financial attributes may dominate numerical features, remains underexplored. The present work bridges that gap by demonstrating empirically that embedding-based representations of categorical bond attributes outperform classical approaches (one-hot and numerical baselines) in sparse-issuer regimes and achieve comparable or superior performance to supervised metric learners such as random forests, particularly when data sparsity is pronounced.

\section{Representation Learning} \label{sec:model}
The core challenge in financial bond similarity search lies in constructing a representation of bonds that captures the complex relationships among their attributes, market conditions, and liquidity characteristics. Traditional methods either rely on predefined distance metrics or extract similarity from specific model architectures, but they often fail to generalize across diverse market regimes or adequately encode the nuanced interactions among features. Representation learning provides a principled approach to address these challenges by automatically learning embeddings of textual and categorical data into a latent space, where proximity reflects meaningful similarity \parencite{replearning}. In particular, for high-cardinality categorical variables (such as bond issuer identity, with millions of possibilities across global markets, or granular industry classification), one-hot encoding suffers from the curse of dimensionality: it produces sparse, orthogonal representations that prevent the model from learning meaningful similarities between categories. Learned embeddings, by contrast, map categories into a continuous latent space where semantic proximity is captured by vector distance. This enables superior generalization in sparse-data regimes, where the model must predict similarity for issuer or sector combinations not extensively represented in training data.

With the growth of unstructured financial data (such as issuer disclosures, news sentiment, and transcripts) deep learning architectures provide opportunities to learn richer representations. Autoencoders, variational autoencoders, and graph neural networks (GNNs) have been proposed to extract embeddings that preserve both numerical and relational information. For corporate bonds, where issuers and securities are embedded in complex credit networks, GNNs can capture structural relationships (e.g., shared sector exposure or counterparty risk) that are overlooked by purely tabular models. Similarly, transformers trained on textual data can produce embeddings that complement traditional quantitative features, enabling similarity search.

Transformer-based embedding models, such as BERT (Bidirectional Encoder Representations from Transformers) \parencite{bert} and sentence BERT (sBERT) \parencite{sbert}, are designed to understand the context of words within a sentence by processing the input text bidirectionally, meaning they consider both the preceding and following words to capture deeper semantic meaning. This bidirectional approach allows transformers to go beyond simple word associations and capture more nuanced aspects of language, such as word sense disambiguation (WSD) \parencite{WDS, transformer}. When applied to embedding categorical financial data, a transformer model generates dense vector representations that capture not only syntactic meaning but also the intent, context, and relationships between financial entities \parencite{WDR}. For example, when embedding bond issuer sectors (Technology, Retail, Manufacturing, etc.), transformers learn to recognize that Technology and E-Commerce are semantically closer than Technology and Banking, a distinction reflecting sector risk correlation and market structure. While BERT processes individual tokens and learns contextual embeddings for each word in a sentence, sBERT is specifically optimized to generate embeddings for entire sentences or documents. sBERT, which builds upon BERT's architecture, is fine-tuned to produce fixed-size dense vectors that represent the overall meaning of a sentence or document, making them ideal for tasks such as semantic similarity, clustering, and information retrieval \parencite{sbert, distilbert}. In the financial domain, contextual embeddings have demonstrated effectiveness in capturing domain-specific semantics of issuer sectors, credit ratings, and market structures.

Pretrained language models such as BERT are trained on large, general-domain corpora (e.g., Wikipedia, BookCorpus) and capture broad linguistic and semantic patterns. However, the financial domain exhibits distinctive characteristics: issuers cluster within sectors and geographies; credit ratings carry specific risk implications; and institutional conventions shape how bonds are classified. Domain-specific fine-tuning adapts the pretrained model's learned representations to these financial nuances, resulting in embeddings where semantic proximity more faithfully reflects economically meaningful relationships \parencite{Gururangan2020DontSP, riskembed}. This is particularly important for bond similarity search, where the embedding model must recognize subtle but economically crucial relationships (e.g., that a Canadian provincial issuer is more similar to other provincial issuers regardless of sector, than to a numerically similar corporate bond from another country). Fine-tuned embeddings capture such domain-specific hierarchies that purely general-purpose models would miss, delivering superior performance on financial downstream tasks.

This section has reviewed the theoretical foundations and practical advantages of representation learning for categorical variables. The embedding-based approach proposed in this work, where similarity is defined by learned semantic relationships among categories, is deliberately orthogonal to prediction-based supervised similarity learning methods reviewed in Section 2. Where supervised approaches (random forests, QCML) extract similarity from the structure of a trained model or learned quantum geometry optimized for a prediction task (e.g., yield prediction), embeddings capture similarity defined by the inherent semantic structure and relationships among categories themselves, independent of any specific prediction target. This is a fundamental distinction: embeddings ask "How similar are issuer A and issuer B in conceptual or semantic space?", while supervised methods ask "Given that we wish to predict yield, which bonds' yields should influence the prediction for a test bond?". Both are valid notions of similarity, and they need not align. The comparative strengths of embedding-based and prediction-based approaches are empirically evaluated in Sections 5--6, where we demonstrate that embedding-based representations of non-financial categorical attributes outperform one-hot encodings and achieve comparable or superior performance to tree-based methods like random forests, particularly in sparse data regimes where semantic similarity becomes especially valuable. Future work may investigate hybrid models that integrate learned categorical embeddings with supervised metric learning frameworks, exploiting complementary strengths in both categorical representation and numerical feature dynamics.

\section{Data Description} \label{sec:data}
For this study, we construct our dataset using multiple years of daily bond data obtained from Bloomberg, covering a broad cross-section of issuers and sectors. The dataset consists of thousands of individual securities, each observed over time and characterized by more than 200 raw attributes, including pricing, liquidity, structural, and issuer-specific variables.

Given the high dimensionality of the raw data, not all attributes are equally informative for the task of similarity search. Many features are redundant, noisy, or weakly correlated with target measures such as yield, spread, or liquidity. To focus the analysis and reduce dimensionality, we conducted a systematic feature selection process that combined domain expertise with empirical analysis of variable stability and predictive value.

As a result, we hand-selected six core features that capture the most relevant aspects of bond similarity. All six of these features are categorical variables, which allows us to group bonds along discrete economic or structural dimensions that are meaningful to traders and portfolio managers. The selected features are outlined in Table~\ref{tab:features}. These core variables are non-financial in the sense that they do not directly encode prices, spreads, or cash-flow magnitudes. They characterize issuer identity, sector, and geographic footprint, which the empirical analysis shows to be highly informative. In addition, the categorical nature of these features is particularly well suited to representation learning approaches. Traditional distance metrics often struggle with high-cardinality categorical variables (e.g., issuer or industry), as they lack a natural ordering or scale. By embedding these categories into a latent representation space, similarity learning models can uncover meaningful relationships (e.g., between industries with correlated risk, or between issuers within the same rating tier) that are not explicit in the raw data.

Although this study narrows its focus to six categorical features for clarity and interpretability, the framework is extensible. Future work may reincorporate additional attributes from the broader Bloomberg dataset. Subsequent experiments explicitly contrast representations based on these categorical attributes with one-hot encodings and with models that rely primarily on numerical financial variables, in order to quantify their relative importance for spread curve reconstruction.

\begin{table}[t]
\centering
\caption{Hand-selected categorical features for bond similarity learning.}
\label{tab:features}
\begin{tabular}{@{}lll@{}}
\toprule
\textbf{Feature} & \textbf{Category} & \textbf{Description} \\ 
\midrule
Issuer Industry & Issuer Attribute & Broad sector classification of the issuing firm (e.g., \textit{Industrial}). \\
Market Issue Type & Market Attribute & Market scope of the bond (e.g., \textit{Global}). \\ 
Industry Group & Market Attribute & Intermediate sector grouping (e.g., \textit{Computers}). \\ 
Industry Subgroup & Market Attribute & Fine-grained industry category (e.g., \textit{E-Commerce}). \\ 
Country of Domicile & Geographic Attribute & Issuer's country of registration or operation (e.g., \textit{US}). \\ 
Issuer Identity & Issuer Attribute & Categorical identifier of the issuing firm (e.g., \textit{APPLE}). \\ 
\bottomrule
\end{tabular}
\end{table}

\section{Methodology} \label{sec:methods}
Our methodology integrates \textit{representation learning} with \textit{post-processing filters} to construct a practical and robust framework for bond similarity search. The process consists of two main components: (1) generating similarity scores via a fine-tuned, pretrained embedding model, and (2) applying post-filters to refine the ranked list of candidate bonds. Finally, we evaluate our methodology through a structured experimental setup that measures how effectively the augmented bond catalog can reproduce realistic bond spread curves compared to the actual market data (see Section~\ref{subsec:evaluation}).

\subsection{Embedding}
\begin{figure}[t]
    \centering
    \includegraphics[width=0.8\textwidth]{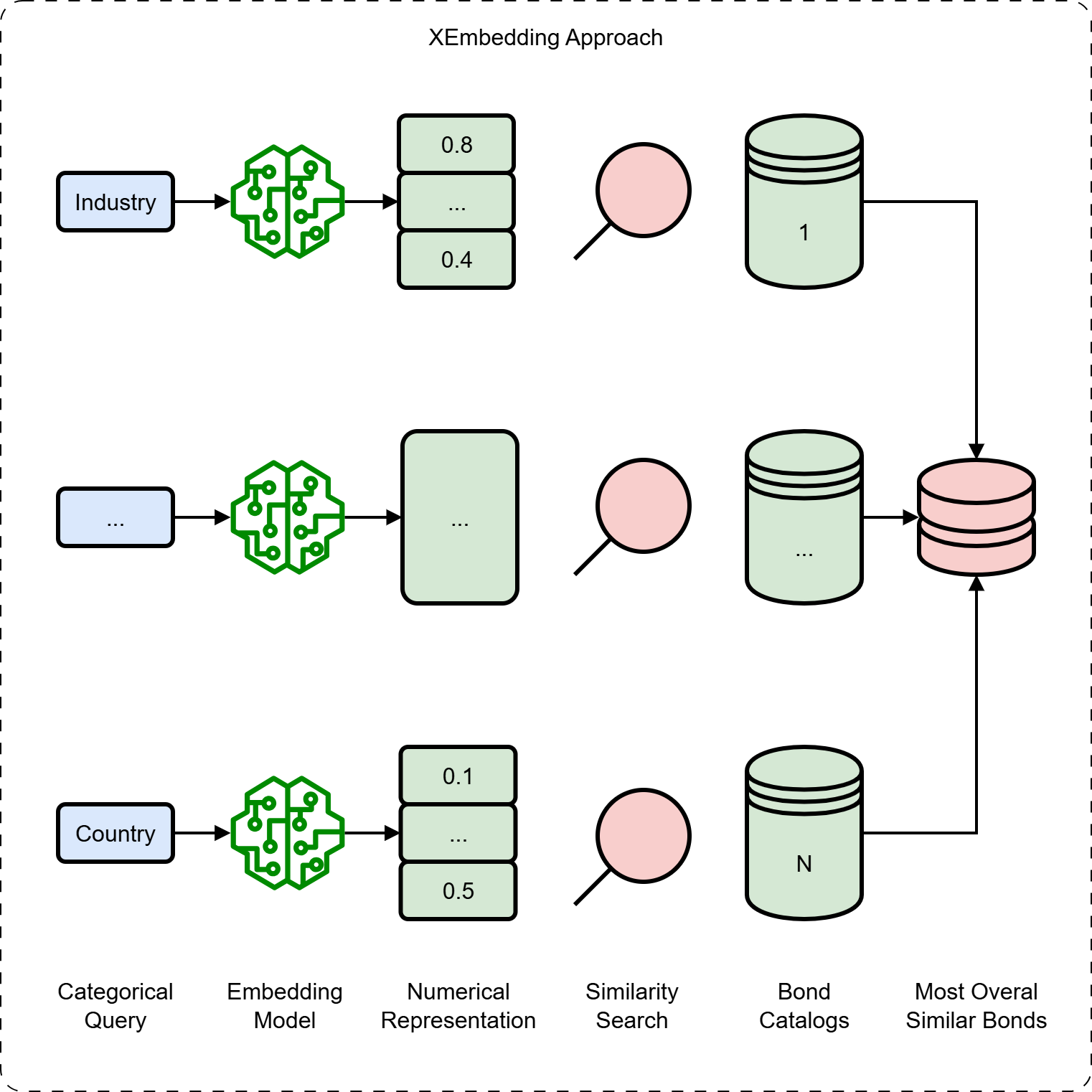}
    \caption{Overview of the embedding-based approach for bond similarity search. 
    Categorical queries (e.g., Industry, Country) are passed through a fine-tuned 
    embedding model to produce dense numerical representations. These embeddings 
    are then compared via similarity search against bond catalogs. After aggregating 
    results across all features, the system outputs the most overall similar bonds.}
    \label{fig:xembedding}
\end{figure}

We begin with a pretrained embedding model that has been fine-tuned on a broad range of financial data \parencite{riskembed}. Fine-tuning enables the model to learn domain-specific representations in which bonds with similar categorical profiles are embedded close to one another in latent space. Then, each bond $b_i$ with feature vector $x_i$ is mapped into an embedding vector $z_i \in \mathbb{R}^d$, where $d$ is the dimension of the learned latent space. Similarity between two bonds $b_i$ and $b_j$ is computed as:  

\begin{equation}
\textnormal{sim}(b_i, b_j) = \frac{\langle z_i, z_j \rangle}{\|z_i\| \cdot \|z_j\|},
\end{equation}

i.e., the cosine similarity between their embedding vectors. For each query bond, the model produces a \textit{ranked list of nearest neighbors} by sorting other bonds in descending order of similarity. As illustrated in Figure~\ref{fig:xembedding}, this approach leverages the advantages of representation learning: categorical features are embedded into continuous vectors that preserve semantic relationships (e.g., issuers in related industries). Unlike handcrafted distance metrics, the embedding model adaptively captures these relationships during training.  

\begin{figure}[t]
    \centering
    \includegraphics[width=0.8\textwidth]{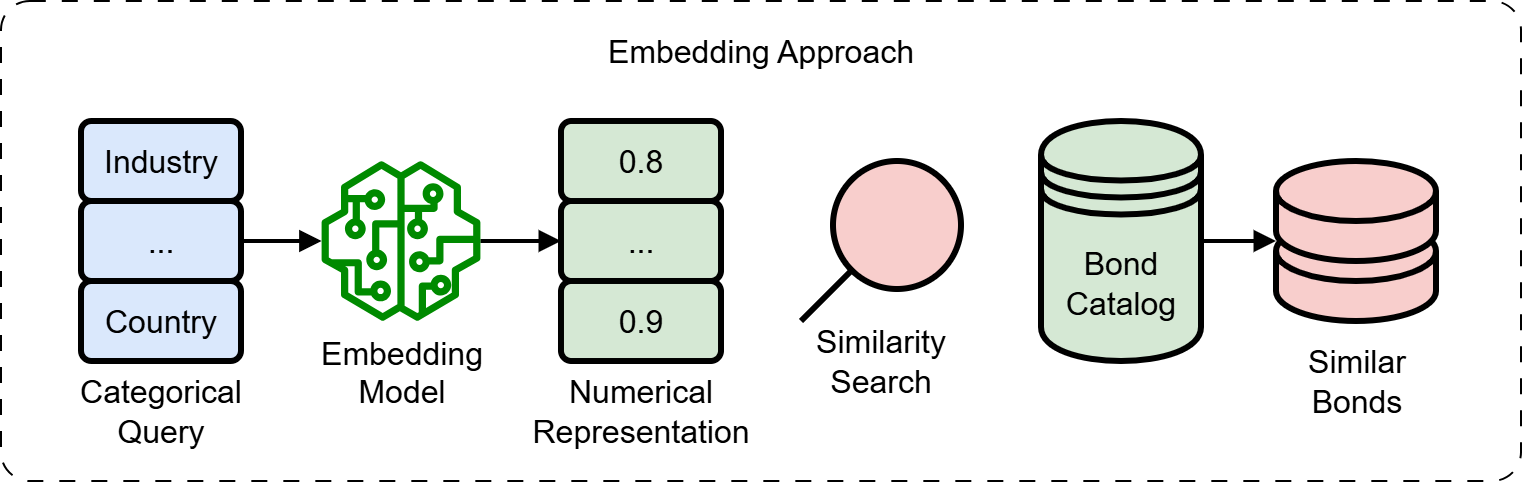}
    \caption{Initial embedding-based methodology for bond similarity search. In this design, all categorical features are concatenated and jointly passed through the embedding model to produce dense numerical representations. Although this approach captures feature interactions more effectively, it offers less interpretability compared to the revised, feature-wise embedding approach.}
    \label{fig:embedding}
\end{figure}

It is worth mentioning that we initially proposed an embedding-based methodology in which all categorical bond features (e.g., industry, rating, and country) were concatenated into a single textual sequence and jointly passed through the embedding model (as illustrated in Figure~\ref{fig:embedding}). While this formulation captured complex interactions among features and resulted in slightly higher similarity accuracy, it suffered from reduced interpretability and made it difficult to isolate the contribution of each feature to the final similarity score. To enhance explainability, we revised the design to the current approach, where each categorical feature is independently embedded before the similarity search and aggregation stages. This modification provides greater transparency by allowing analysts to examine feature-level similarities across bonds. However, because feature interactions are no longer jointly modeled, the overall similarity accuracy experiences a minor reduction, reflecting a deliberate trade-off between model performance and interpretability. The categorical embeddings derived from this model are referred to as XEmbedding.

\subsection{Post-Filtering}
While embeddings provide a powerful foundation for similarity search, additional post-filters could be utilized to ensure that results satisfy practical trading and portfolio management constraints. After generating the sorted list of candidate bonds from the embedding model, we apply the following sequence of filters to refine the output:  

\begin{enumerate} 
    \item \textbf{Currency Filter} -- Retain only bonds denominated in the same currency as the query bond.  
    \item \textbf{Maturity Filter} -- Restrict candidates to those within a defined lower- and/or upper-bound time to maturity.  
    \item \textbf{Rating Filter} -- Maintain credit quality alignment by allowing only bonds within a predefined rating tolerance.  
\end{enumerate}

These filters are applied progressively, narrowing down the candidate set. The final output is a refined, ranked list of bonds that balances \textit{semantic similarity in the embedding space} with \textit{practical market constraints}.

\subsection{Evaluation} \label{subsec:evaluation}
\begin{figure}[!t]
    \centering
    \includegraphics[width=\linewidth]{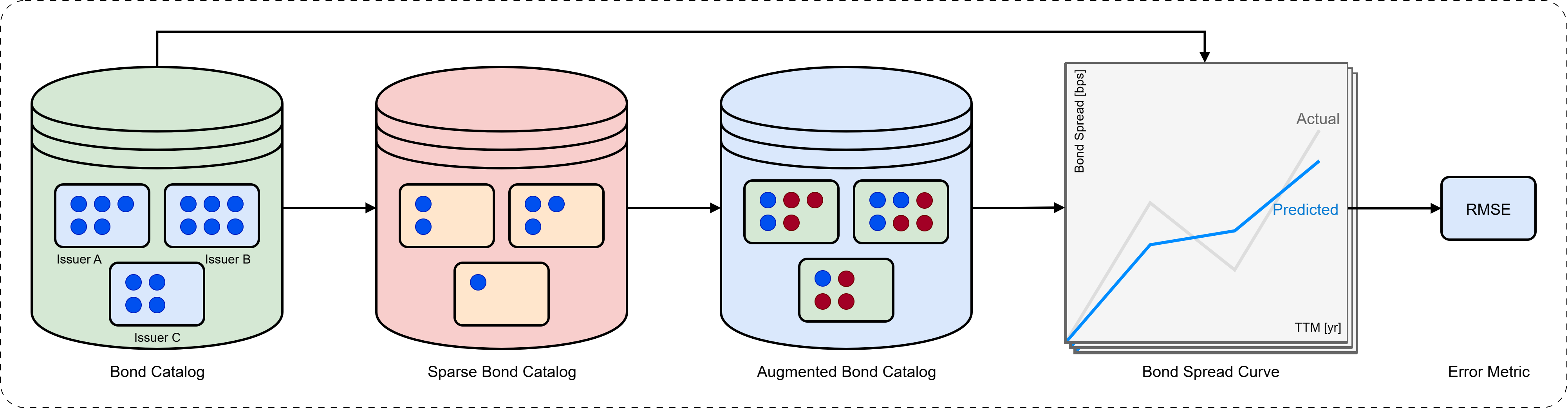}
    \caption{Evaluation pipeline for the proposed bond similarity search framework. Starting from a complete bond catalog, a subset of non-sparse issuers is selected. A fixed number of bonds are randomly removed to create sparsity. The sparse issuers are then augmented using the proposed similarity-based methodology. The augmented catalog is used to fit predicted bond spread curves via the NS model, which are compared to the actual curves using the RMSE error metric.}
    \label{fig:evaluation}
\end{figure}

To evaluate the effectiveness of the proposed bond similarity search and augmentation framework, we design an experiment based on the process illustrated in Figure~\ref{fig:evaluation}. The objective is to assess how well the augmented bond catalog can reproduce realistic bond spread curves compared to the actual market data.

We begin with the \textit{bond catalog}, which contains multiple issuers with varying bond densities. Some issuers have sufficient bonds across different maturities (non-sparse issuers), while others have only a few outstanding bonds (sparse issuers). For evaluation purposes, we first select a subset of non-sparse issuers, ensuring that each selected issuer has a sufficiently rich term structure to serve as ground truth. From each selected issuer, we randomly drop a fixed number of bonds to artificially create sparsity in their term structures. These reduced sets form the \textit{sparse bond catalog}, mimicking real-world cases where certain issuers have limited bond data available.

Next, we apply our proposed augmentation methodology to these sparse issuers. The similarity-based framework identifies comparable bonds from other issuers in the catalog and augments the sparse issuer's data with similar bonds. The resulting \textit{augmented bond catalog} represents an enhanced dataset where each issuer's term structure is partially reconstructed using embedding-based similarity retrieval. Once the augmented dataset is generated, we use a Nelson--Siegel (NS) model to fit the \textit{predicted bond spread curve} for each issuer. We then compare these predicted curves against the \textit{actual bond spread curves} derived from the original (non-sparse) catalog. The discrepancy between the predicted and actual spread curves is quantified using the Root Mean Square Error (RMSE) metric:

\begin{equation}
\textnormal{RMSE} = \sqrt{\frac{1}{N} \sum_{i=1}^{N} (\hat{y}_i - y_i)^2}
\label{eq:rmse}
\end{equation}

where $\hat{y}_i$ and $y_i$ denote the predicted and actual bond spreads, respectively. A lower RMSE indicates that the augmented issuer more accurately reproduces the true bond spread dynamics, reflecting the quality of the similarity-based augmentation.

\subsection{Benchmarking} \label{subsec:benchmark}
To rigorously assess the value of embedding-based representations for categorical non-financial bond attributes, our model is benchmarked against a one-hot encoded baseline that represents the conventional approach to handling categorical variables in financial similarity search. Both models operate within the same experimental framework of sparse-issuer augmentation followed by Nelson--Siegel spread-curve fitting, ensuring a controlled comparison of representation quality while holding all other aspects including the data universe, sparsity protocol, post-filters, and evaluation metric constant.

The one-hot baseline encodes each of the six categorical features (issuer industry, market issue type, industry group, subgroup, country of domicile, issuer identity, currency, and bond rating) as sparse binary vectors. For a given query bond, similarity is computed as a weighted average of exact-match frequencies across features, where weights reflect domain-informed priorities (e.g., issuer industry and rating receive higher weights than market issue type). The top-K nearest neighbors are then selected using this aggregate score, identical to the XEmbedding pipeline, and subjected to the same post-processing filters. These neighbors augment the sparse issuer catalog, and the NS model is fit to the augmented term structure to predict spreads across maturities.

This setup isolates the effect of categorical representation: one-hot assumes orthogonality between categories and relies on exact matches or simple aggregation, while our model captures semantic proximity through dense learned embeddings. This methodology quantifies whether nuanced categorical similarity (XEmbedding) meaningfully improves spread curve reconstruction over rigid one-hot encodings, directly testing the hypothesis that non-financial attribute representation dominates numerical financial detail in sparse-data regimes.

\section{Results and Discussion} \label{sec:results}
In this section, the proposed similarity search methodology is applied to a large universe of corporate bonds (i.e., 2,500 bonds from 250 unique issuers) to identify instruments that share the most comparable characteristics with a given query bond. Furthermore, to assess the effectiveness of the proposed bond similarity search and augmentation methodology, we conduct a series of experiments comparing the predicted bond spread curves against the actual market data. The evaluation is based on a set of issuers with varying levels of bond sparsity, mimicking real-world scenarios where some issuers have sparse bond data while others have more comprehensive term structures. A more extensive set of example results is provided in the Appendix.

\subsection{Similarity Search} \label{subsec:similarity}
\begin{figure}[!t]
    \centering
    \includegraphics[width=0.5\paperheight,height=\paperwidth,keepaspectratio,angle=90]{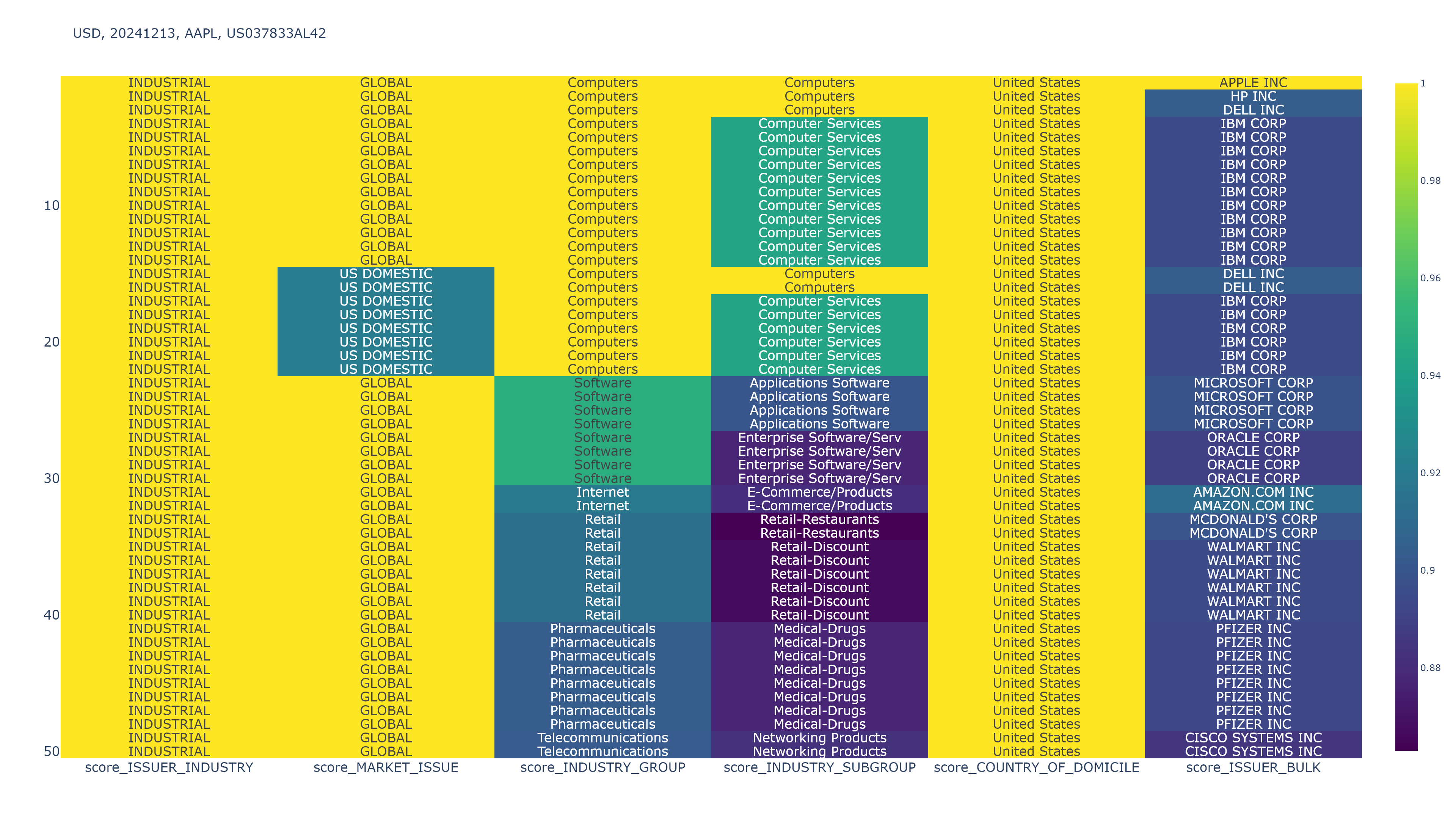}
    \includegraphics[width=0.5\paperheight,height=\paperwidth,keepaspectratio,angle=90]{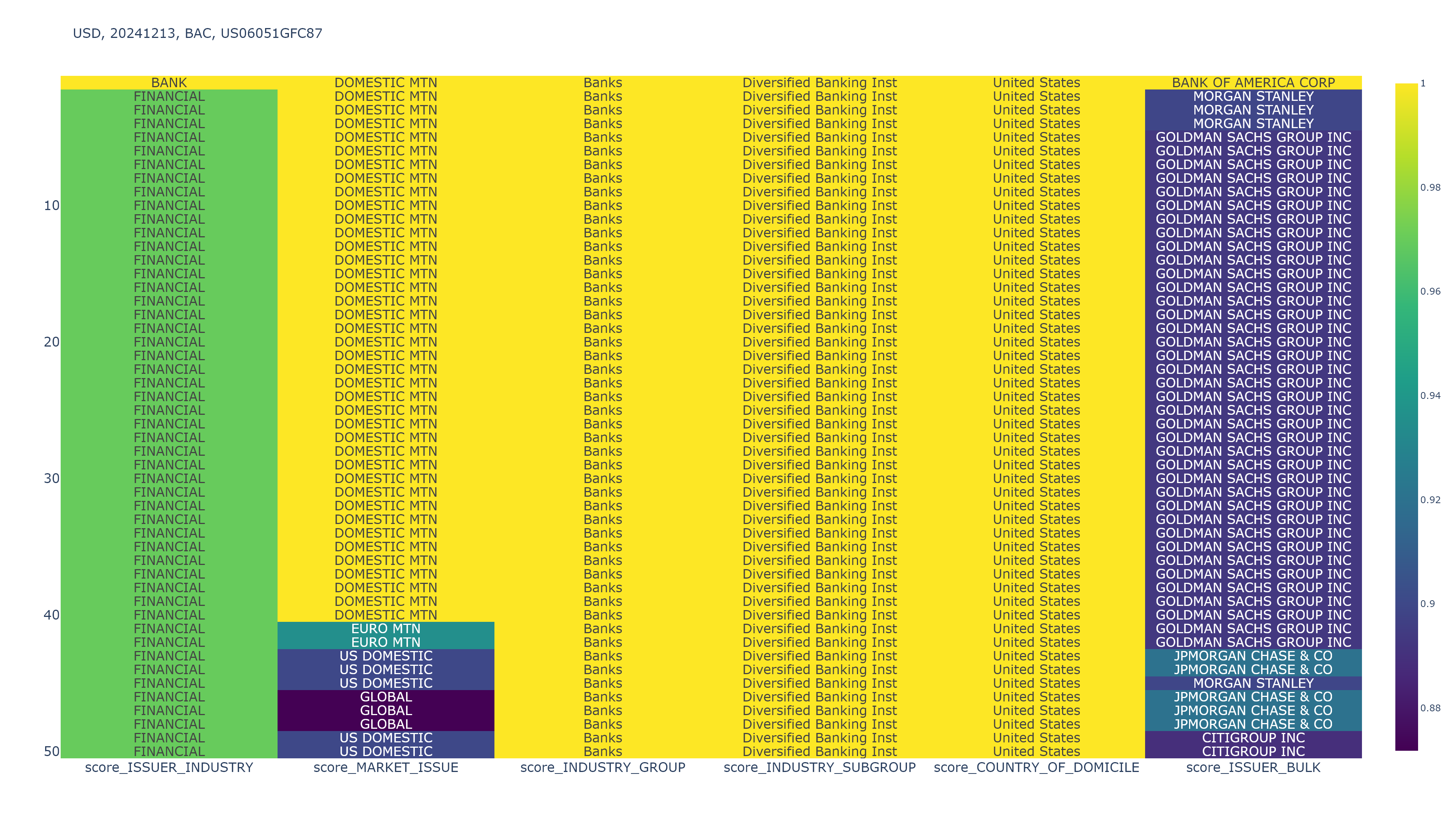}
    \caption{Visualization of bond similarity search results for the query bonds of AAPL US037833AL42 (left) and BAC US06051GFC87 (right). The first row represents the query bond's profile, while subsequent rows show the most similar bonds ranked by cosine similarity in embedding space. The color intensity indicates similarity, with higher scores (topmost column) reflecting closer structural and economic resemblance to the query bond.}
    \label{fig:aapl1bac1}
\end{figure}

The similarity measure is derived using an embedding approach that employs categorical attributes of each bond. Specifically, categorical features such as issuer industry and industry group were transformed into dense vector representations through an embedding model. For instance, the visualization shown in Figure~\ref{fig:aapl1bac1} demonstrates how the method retrieves bonds that are structurally and economically similar to the query bonds from Apple Inc. (AAPL) and Bank of America Corp. (BAC) (issued on 2024-12-13). The heatmap presents the categorical profile of the query and its nearest neighbors, ranked by their similarity scores, which are shown along the right-hand side. The first row represents the query bond itself, serving as the reference point with a similarity score of 1.00. Subsequent rows correspond to bonds with progressively lower scores, indicating decreasing levels of similarity. The retrieved bonds cluster around specific industry and issuer types consistent with the market characteristics of Apple and Bank of America Corp. The top-ranked matches include bonds and issuers that share the similar global industrial profiles and operate within similar subsectors. This demonstrates the model's capacity to discern economic comparability within the broader industrial and consumer sectors, rather than merely matching by superficial categorical overlap. Further down the rankings, the model surfaces bonds issued by companies that, while belonging to different subsectors, remain within the same global corporate class and exhibit similar credit quality. This reflects the embedding model's ability to capture second-order similarity: bonds that differ in industry but align in credit and structural attributes. The gradual decline in similarity scores illustrates a smooth transition from highly comparable peers to tangentially related securities, confirming the embedding space's continuous structure.

Moreover, to visualize embedding quality, we construct two-dimensional projections of the embedding space for three key features (industry subgroup, issuer bulk, and country of domicile) shown in Figure~\ref{fig:2demb}. Points are colored by their similarity to a reference category in the original high-dimensional space (i.e., with 768 dimensions); the smooth color gradients in the projections indicate that the 2D layout remains faithful to the underlying categorical similarity structure. Across all three views, entities with similar categorical attributes exhibit clear clustering patterns, indicating that the embedding space captures meaningful semantic and structural relationships. In particular, separation by issuer and geographic domicile suggests that both firm-specific and regional signals are strongly encoded, while the industry-level view reflects finer-grained similarities within related sectors. Together, these projections provide qualitative evidence that the learned representations preserve relevant domain structure beyond the original feature space.

Overall, these results validate the robustness of the similarity search methodology. By combining categorical embeddings with post-filtering constraints, the model successfully prioritizes economically relevant comparables while maintaining diversity in issuer types and industries. The visualization provides intuitive interpretability, allowing researchers and practitioners to observe both the alignment of categorical attributes and the gradient of similarity scores across potential matches. This approach offers significant utility in fixed-income research applications such as relative value analysis, liquidity estimation, and risk benchmarking, where identifying functionally similar bonds is essential.

\subsection{Bond Spread Curve} \label{subsec:curve}
\begin{figure}[!t]
    \centering
    \includegraphics[width=0.49\linewidth]{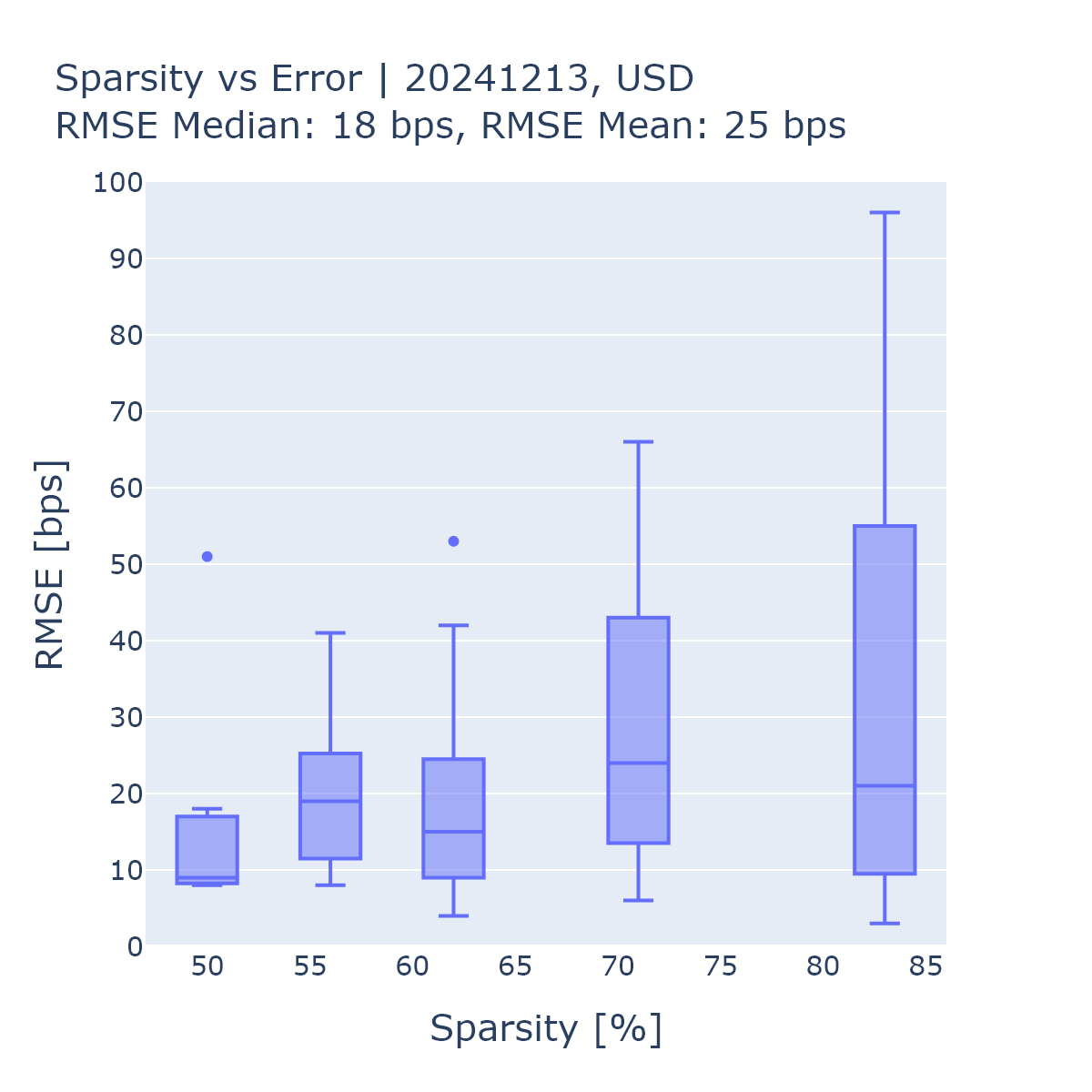}
    \includegraphics[width=0.49\linewidth]{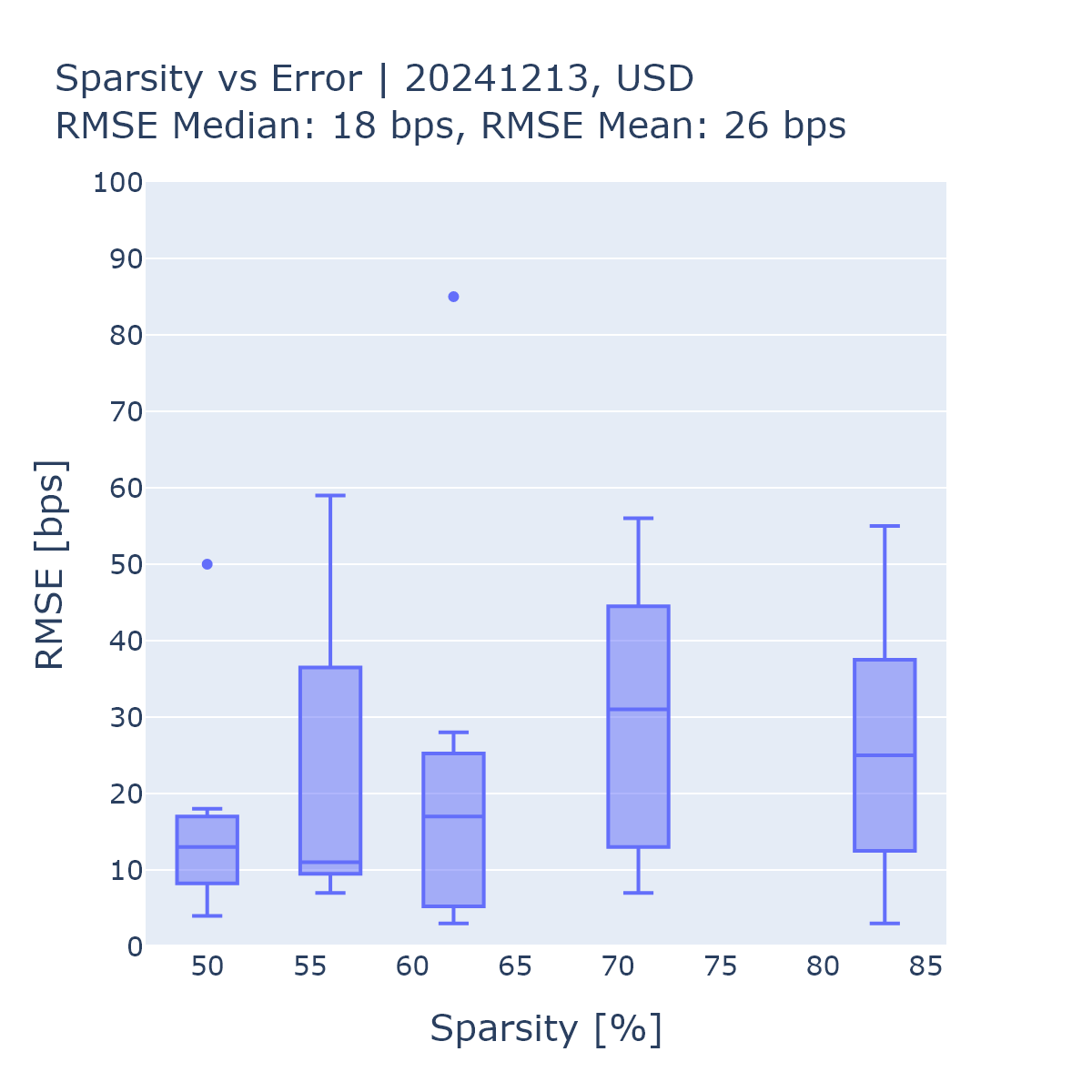}
    \caption{Comparison of overall performance of our model without (left) and with (right) post-filters. Post-filtering improves model stability under sparsity by tightening error distributions and reducing tail risk.}
    \label{fig:errors}
\end{figure}

The results in this subsection evaluate the effectiveness of our model through CDS spread-curve reconstruction, using the Nelson--Siegel (NS) model fit to augmented sparse-issuer catalogs. Figure~\ref{fig:errors} compares the overall performance of our model without (left) and with (right) post-filters. Both panels plot RMSE against sparsity levels (where sparsity is defined as $\#\textnormal{Queries} / (\#\textnormal{Queries}+\#\textnormal{Similars})$) for multiple issuers, using boxplots to summarize the distribution of errors at each sparsity bin alongside individual issuer scatter points. The x-axis spans increasing sparsity from left to right, while the y-axis captures RMSE in basis points, revealing the expected positive trend where typically higher sparsity correlates with elevated prediction errors due to fewer query bonds available for augmentation.

The left panel shows the model results without post-filtering, achieving a median RMSE of 18 bps and mean of 26 bps across issuers driven by extreme outliers exceeding 90 bps that widen the whiskers dramatically. In contrast, the right panel shows the post-filtered model achieving a comparable median of 18 bps but a lower mean RMSE of 25 bps. Post-filtering further tightens the error distribution and preserves strong clustering even under high sparsity (70--85\%). Overall, beyond illustrating aggregate model performance, this comparison highlights how post-filtering mitigates sensitivity to poor categorical matches in sparse regimes, leading to improved stability evidenced by narrower boxplots, reduced tail risk, and more consistent performance.

\begin{figure}[!t]
    \centering
    \includegraphics[width=0.49\linewidth]{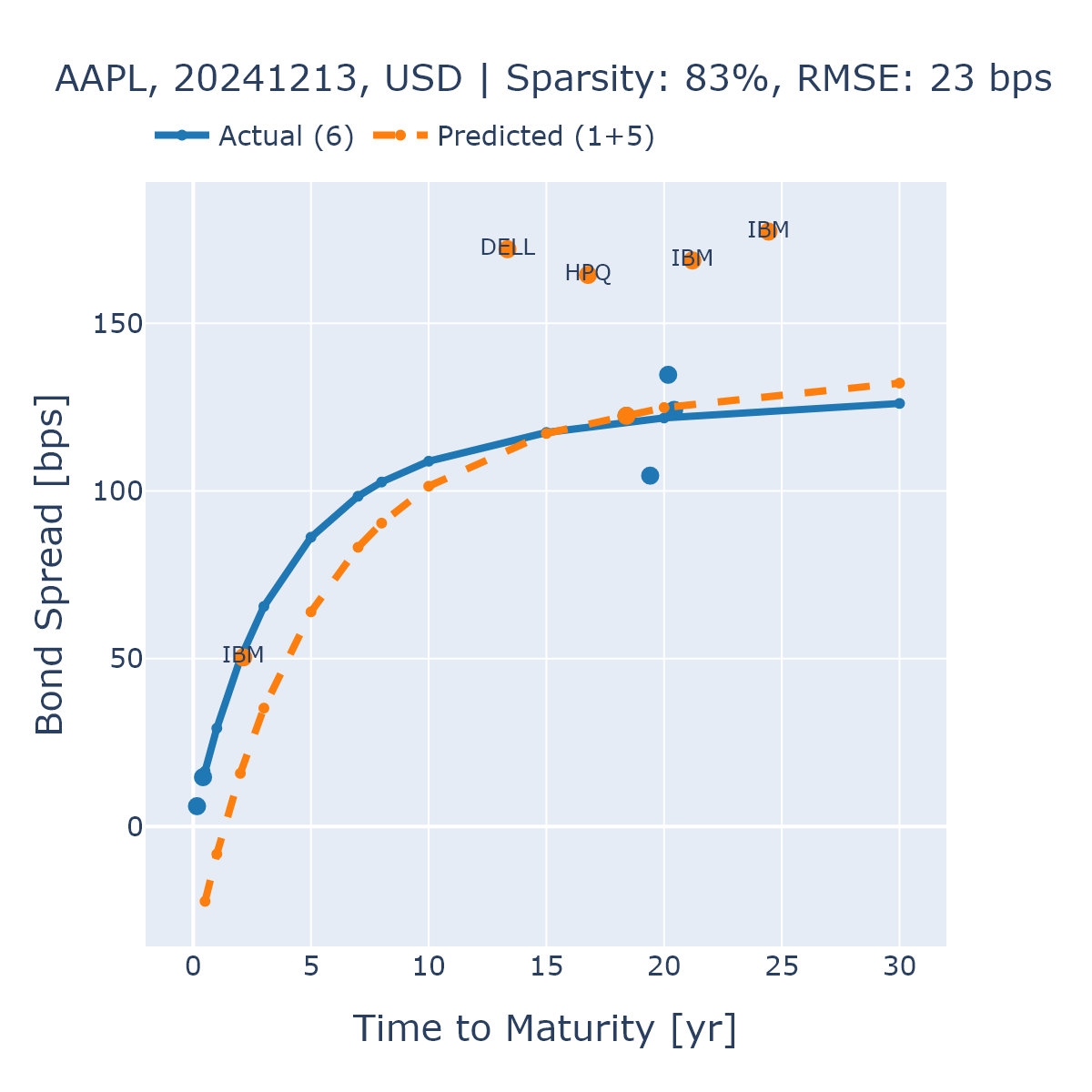}
    \includegraphics[width=0.49\linewidth]{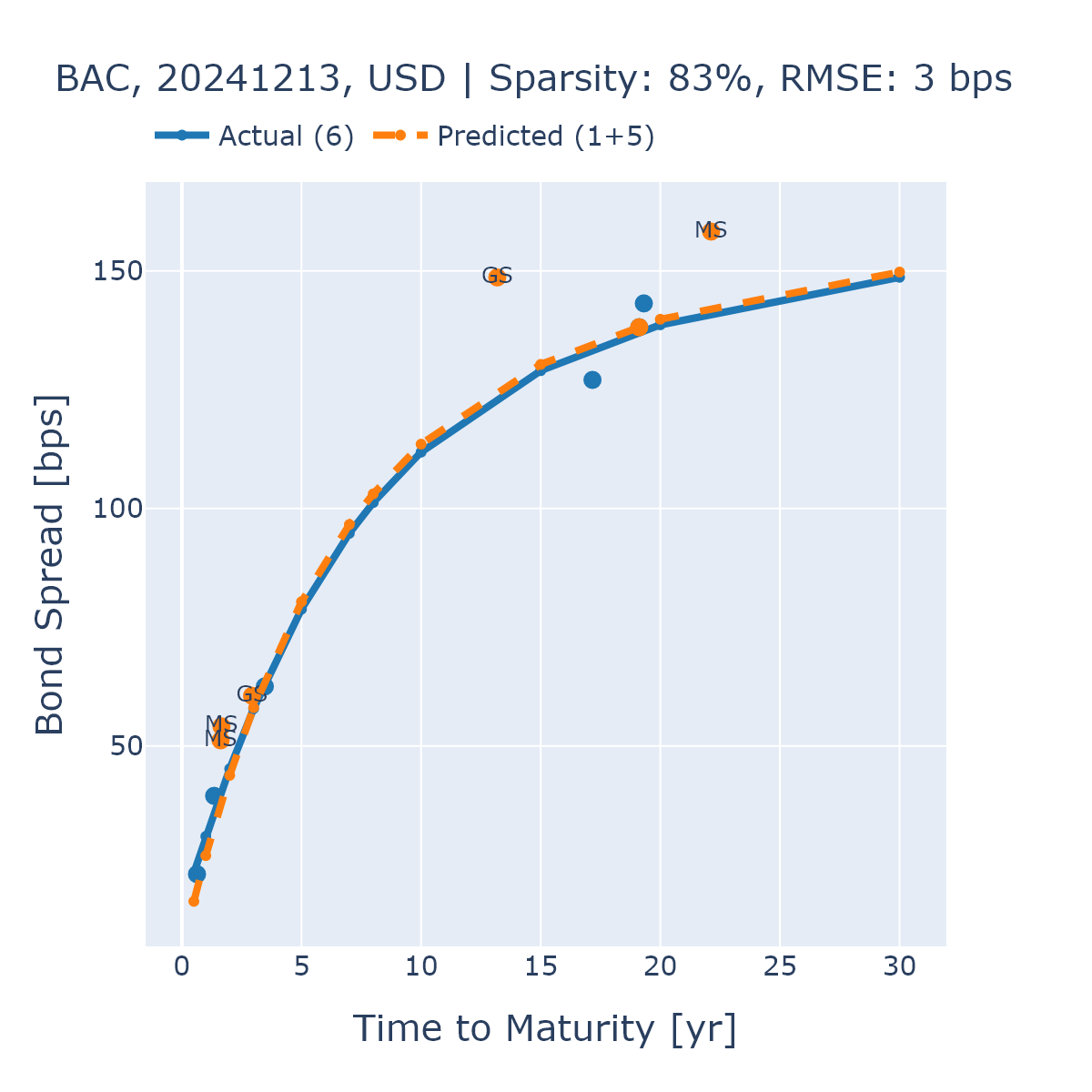}
    \caption{Comparison of predicted and actual CDS spread curves for AAPL (left) and BAC (right) bonds on 2024-12-13 under a high sparsity regime (83\%). Blue markers represent the actual bonds, non-annotated orange markers show predictions for the query, and annotated orange markers indicate specific predicted bonds (e.g., DELL, IBM, and GS). Despite limited information, the model captures the overall curve shape, achieving RMSEs of 23 bps for AAPL and 3 bps for BAC.}
    \label{fig:aapl0bac0}
\end{figure}

We use Apple Inc. (AAPL) and Bank of America Corp. (BAC) as representative examples to illustrate the performance of our framework for augmenting sparse bond catalogs. Figure~\ref{fig:aapl0bac0} compares actual and predicted bond spread curves as a function of time to maturity for the two issuers (Apple on the left and Bank of America on the right on 2024-12-13) both under a high sparsity regime of 83\%. Blue solid lines represent observed spreads, while orange dashed lines denote model predictions constructed from limited tenor information. In both panels, the model captures the overall term-structure shape well, reproducing the steep increase at short maturities and the gradual flattening at longer horizons. Individual bond observations are overlaid, illustrating how sparse and uneven the available data are across maturities.

Quantitatively, the model achieves an RMSE of 23 bps for Apple and a much lower RMSE of 3 bps for Bank of America, indicating issuer-dependent performance under identical sparsity conditions. For Apple, larger deviations appear at longer maturities, where sparse categorical matches lead to noticeable but controlled prediction errors. In contrast, the BAC curve shows near-perfect alignment between actual and predicted spreads across the full maturity range, demonstrating strong robustness even with limited data. Together, these examples highlight that while high sparsity can expose sensitivity to issuer-specific structure, the model generally preserves curve shape and delivers stable predictions, with performance improving when categorical alignment is strong.

The results demonstrate that our bond similarity search and augmentation framework is capable of effectively reconstructing bond spread curves, with a relatively low RMSE and good performance in cases of significant data sparsity. This suggests that the approach can be applied to enhance datasets with missing or sparse bond data, providing more realistic bond term structures for credit-risk modeling and investment decisions. Additional qualitative and quantitative results across a broader set of issuers are provided in Appendix~\ref{apndx:search}.

Additionally, we benchmark our model (the "XEmbedding" model) against a baseline that encodes categorical features using one-hot representations (the ``one-hot'' model; with post-filters applied to both). They differ fundamentally in how they represent categorical bond attributes, leading to stark contrasts in spread curve reconstruction accuracy. The one-hot approach encodes features like issuer industry, domicile, and market type as sparse binary vectors, treating categories as orthogonal with no inherent similarity, thus relying on exact matches or simple aggregation rules to select neighbors. In contrast, XEmbedding uses dense transformer-based embeddings to capture semantic and hierarchical relationships, such as placing "Retail--Discount" closer to "E-Commerce" than to "Banks".

\begin{figure}[!t]
    \centering
    \includegraphics[width=0.49\linewidth]{figs/_errorm.png}
    \includegraphics[width=0.49\linewidth]{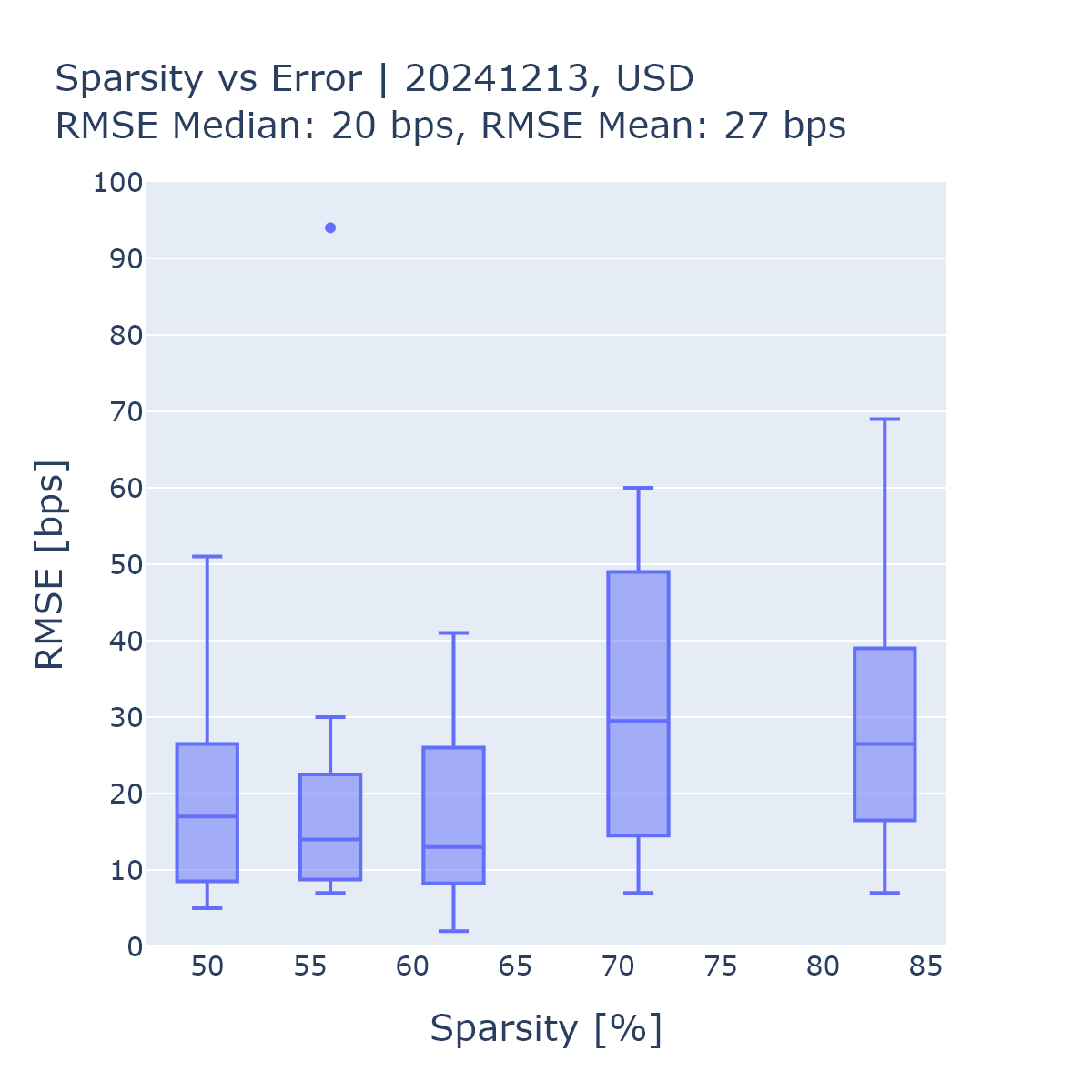}
    \caption{Comparison of overall performance of our model (the "XEmbedding" model) and a baseline that encodes categorical features using one-hot representations (the ``one-hot'' model).}
    \label{fig:errors_onehot_xemb}
\end{figure}

The plots shown in Figure~\ref{fig:errors_onehot_xemb} summarize the aggregate performance of the XEmbedding model (left) versus the one-hot baseline (right) across multiple issuers. The comparison of the two models across varying sparsity levels indicates that our model consistently outperforms the one-hot baseline. The XEmbedding model exhibits lower median and mean RMSE values (18 bps and 26 bps, respectively) compared to the one-hot model (20 bps and 27 bps), reflecting more accurate predictions overall. Additionally, the variability of errors in the XEmbedding model remains smaller, particularly at higher sparsity levels, whereas the one-hot model shows wider interquartile ranges and larger outliers, highlighting its reduced robustness as sparsity increases. These results suggest that the learned embeddings provide a more stable and reliable representation of categorical features than one-hot encoding, especially when the data becomes sparse.

\begin{figure}[!t]
    \centering
    \includegraphics[width=0.49\linewidth]{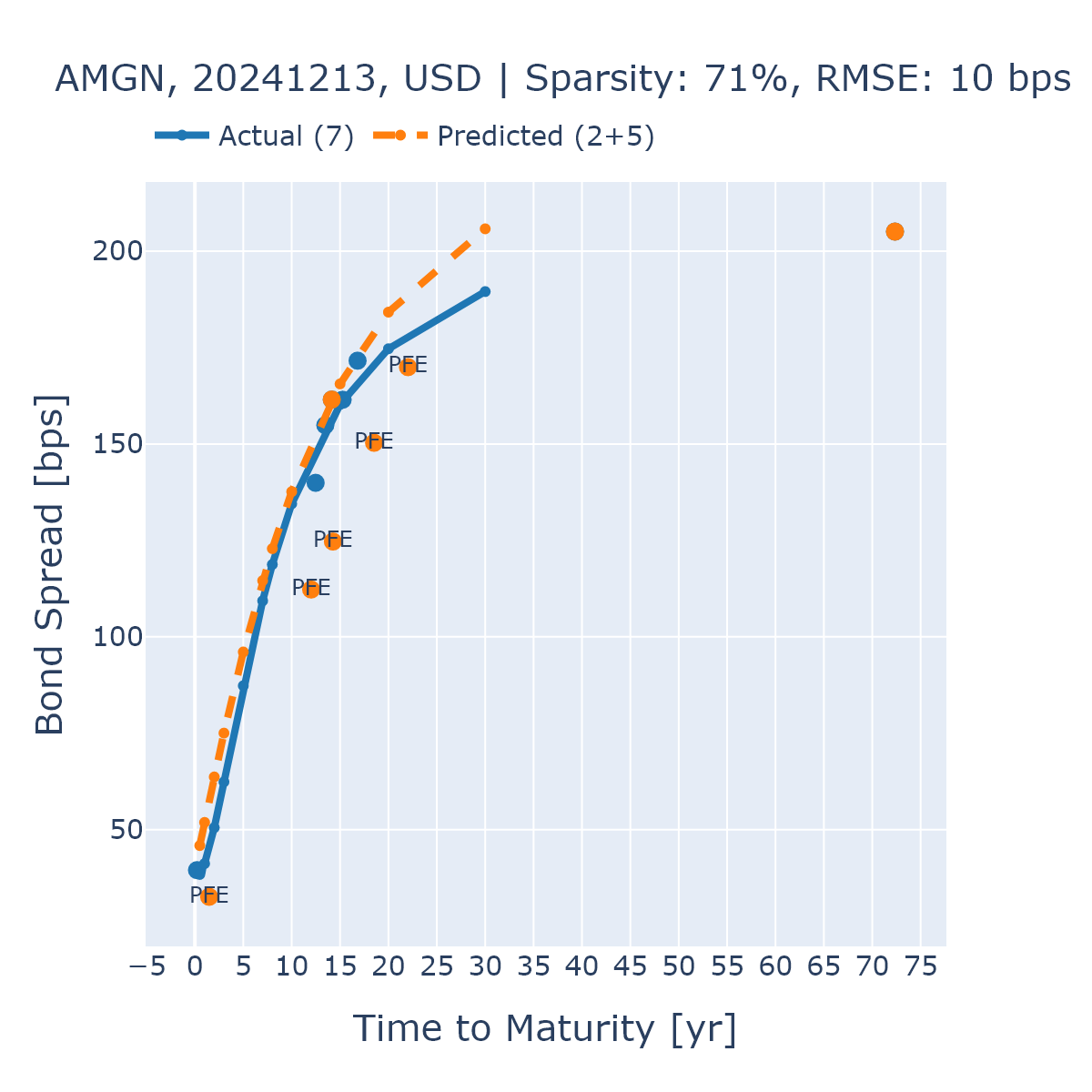}
    \includegraphics[width=0.49\linewidth]{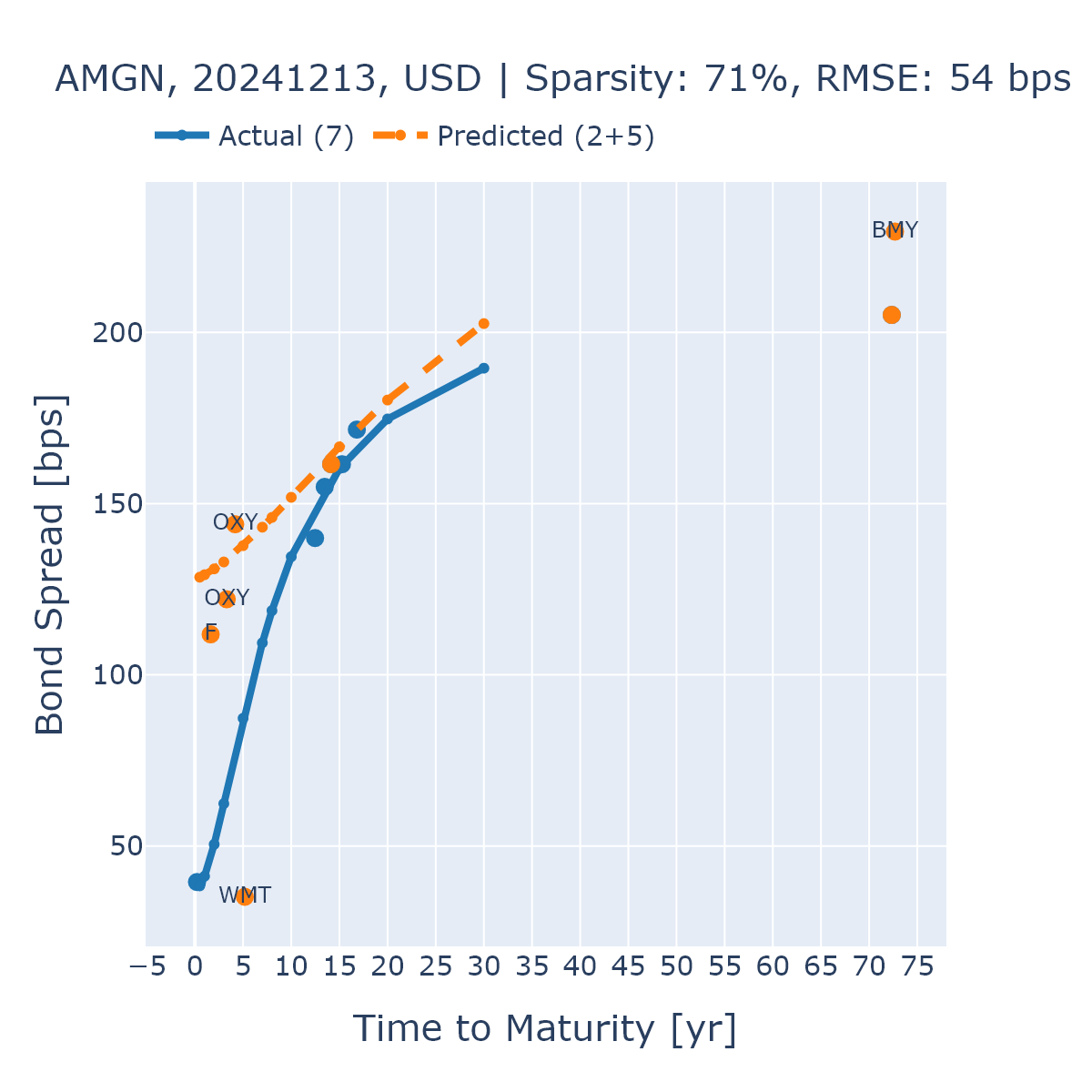}
    \caption{Comparison of XEmbedding (left) and one-hot (right) model prediction versus actual CDS spread curves for AMGN (Amgen Inc.). Annotated orange markers indicate specific predicted bonds, such as PFE (Pfizer Inc.), which belongs to a similar industry group, and OXY (Occidental Petroleum Corp.), which does not.}
    \label{fig:amgn_onehot_xmb}
\end{figure}

As an example, we consider the results of the issuer AMGN (Amgen Inc. is a biopharmaceutical company) shown in Figure~\ref{fig:amgn_onehot_xmb}. It illustrates a concrete issuer-level comparison between XEmbedding and one-hot representations for AMGN under 71\% sparsity, where only 2 query bonds are retained and augmented with 5 similar neighbors. The XEmbedding model (left) produces a predicted curve that closely tracks the actual term structure, achieving an RMSE of 10 bps with smooth alignment across maturities and minimal deviation even at the long end. In stark contrast, the one-hot baseline (right) selects less relevant peers (including Occidental Petroleum Corp. (OXY), Ford Motor Co. (F), and Walmart Inc. (WMT)), resulting in a distorted upward bias and much higher RMSE of 54 bps, as evidenced by the wider spread between predicted and actual points. This example underscores how XEmbedding's nuanced categorical similarities yield more accurate spread reconstructions than rigid exact-match logic, particularly when data is scarce. In fact, the superior performance stems from XEmbedding's ability to quantify partial similarity between non-identical categories, reducing the number of poor matches in high-sparsity regimes. We provide two more examples including KOREA and WFC in Appendix~\ref{apndx:bench}.

\section{Conclusion}
This study presented an embedding-based framework that prioritizes categorical, non-financial bond attributes as the primary drivers of bond similarity. Empirically, embedding representations of these categorical variables outperform one-hot and other baselines in reconstructing spread curves from sparse issuer catalogs, suggesting that how categorical information is encoded matters more than adding further numerical detail. Future work could combine embeddings with supervised similarity learners such as random forest proximities to build hybrid models that exploit both high-quality categorical representations and detailed financial dynamics. In addition, future extensions using multimodal or graph-based embeddings may uncover deeper structural relationships among issuers.

\section*{Acknowledgments}
The authors thank the Risk Management team at TD Bank, in particular Ray Westcott and Jonathan Vitrano, for their collaboration and insightful feedback throughout this research. Their support and commitment to advancing AI-driven analytics in financial risk management were invaluable to this work.

\printbibliography

@article{Gururangan2020DontSP,
  title={Don’t Stop Pretraining: Adapt Language Models to Domains and Tasks},
  author={Suchin Gururangan and Ana Marasovi{\'c} and Swabha Swayamdipta and Kyle Lo and Iz Beltagy and Doug Downey and Noah A. Smith},
  journal={ArXiv},
  year={2020},
  volume={abs/2004.10964},
  url={https://api.semanticscholar.org/CorpusID:216080466}
}

@misc{transformers,
      title={Attention Is All You Need}, 
      author={Ashish Vaswani and Noam Shazeer and Niki Parmar and Jakob Uszkoreit and Llion Jones and Aidan N. Gomez and Lukasz Kaiser and Illia Polosukhin},
      year={2023},
      eprint={1706.03762},
      archivePrefix={arXiv},
      primaryClass={cs.CL},
      url={https://arxiv.org/abs/1706.03762}, 
}

@book{Busemeyer_Bruza_2012, place={Cambridge}, title={Quantum Models of Cognition and Decision}, publisher={Cambridge University Press}, author={Busemeyer, Jerome R. and Bruza, Peter D.}, year={2012}}

@article{Desai,
  author  = {Dhruv Desai and Dhagash Mehta},
  title   = {On Robustness of Mutual Funds Categorization and Distance Metric Learning},
  journal = {The Journal of Financial Data Science},
  year    = {2021},
  pages   = {130 - 150},
  url     = {http://jmlr.org/papers/v17/14-168.html}
}

@article{JMLR:v17:14-168,
  author  = {Lucas Mentch and Giles Hooker},
  title   = {Quantifying Uncertainty in Random Forests via Confidence Intervals and Hypothesis Tests},
  journal = {Journal of Machine Learning Research},
  year    = {2016},
  volume  = {17},
  number  = {26},
  pages   = {1--41},
  url     = {http://jmlr.org/papers/v17/14-168.html}
}

@inproceedings{Breiman2004CONSISTENCYFA,
  title={CONSISTENCY FOR A SIMPLE MODEL OF RANDOM FORESTS},
  author={L. Breiman},
  year={2004},
  url={https://api.semanticscholar.org/CorpusID:123042984}
}

@article{Scornet2015,
   title={Consistency of random forests},
   volume={43},
   ISSN={0090-5364},
   url={http://dx.doi.org/10.1214/15-AOS1321},
   DOI={10.1214/15-aos1321},
   number={4},
   journal={The Annals of Statistics},
   publisher={Institute of Mathematical Statistics},
   author={Scornet, Erwan and Biau, Gérard and Vert, Jean-Philippe},
   year={2015},
}

@book{dl,
  author = {Goodfellow, Ian and Bengio, Yoshua and Courville, Aaron},
  description = {Deep Learning - Ian Goodfellow, Yoshua Bengio, Aaron Courville - Google Books;
Complete MIT book available online in chapters. Book has been recommended to me by several people. I therefore recommend it to you.},
  note = {\url{http://www.deeplearningbook.org}},
  publisher = {MIT Press},
  title = {Deep Learning},
  year = 2016
}

@article{dlfinance,
author = {Heaton, J. B. and Polson, N. G. and Witte, J. H.},
title = {Deep learning for finance: deep portfolios},
journal = {Applied Stochastic Models in Business and Industry},
volume = {33},
number = {1},
pages = {3-12},
doi = {https://doi.org/10.1002/asmb.2209},
url = {https://onlinelibrary.wiley.com/doi/abs/10.1002/asmb.2209},
eprint = {https://onlinelibrary.wiley.com/doi/pdf/10.1002/asmb.2209},
year = {2017}
}

@book{ir,
  author       = {Hang Li},
  title        = {Learning to Rank for Information Retrieval and Natural Language Processing,
                  Second Edition},
  series       = {Synthesis Lectures on Human Language Technologies},
  publisher    = {Morgan {\&} Claypool Publishers},
  year         = {2014},
  url          = {https://doi.org/10.2200/S00607ED2V01Y201410HLT026},
  doi          = {10.2200/S00607ED2V01Y201410HLT026},
  isbn         = {978-3-031-01027-9},
  bibsource    = {dblp computer science bibliography, https://dblp.org}
}

@inproceedings{metriclearning,
  title={Metric Learning : A Survey By},
  author={Brian Kulis},
  year={2013},
  url={https://api.semanticscholar.org/CorpusID:262315341}
}

@ARTICLE{replearning,
  author={Bengio, Yoshua and Courville, Aaron and Vincent, Pascal},
  journal={IEEE Transactions on Pattern Analysis and Machine Intelligence}, 
  title={Representation Learning: A Review and New Perspectives}, 
  year={2013},
  volume={35},
  number={8},
  pages={1798-1828},
  doi={10.1109/TPAMI.2013.50}}

@misc{qcml,
      title={Supervised Similarity for High-Yield Corporate Bonds with Quantum Cognition Machine Learning}, 
      author={Joshua Rosaler and Luca Candelori and Vahagn Kirakosyan and Kharen Musaelian and Ryan Samson and Martin T. Wells and Dhagash Mehta and Stefano Pasquali},
      year={2025},
      eprint={2502.01495},
      archivePrefix={arXiv},
      primaryClass={q-fin.ST},
      url={https://arxiv.org/abs/2502.01495}, 
}

@misc{rf,
      title={Supervised similarity learning for corporate bonds using Random Forest proximities}, 
      author={Jerinsh Jeyapaulraj and Dhruv Desai and Peter Chu and Dhagash Mehta and Stefano Pasquali and Philip Sommer},
      year={2022},
      eprint={2207.04368},
      archivePrefix={arXiv},
      primaryClass={q-fin.CP},
      url={https://arxiv.org/abs/2207.04368}, 
}

@misc{riskembed,
      title={Generative AI Enhanced Financial Risk Management Information Retrieval}, 
      author={Amin Haeri and Jonathan Vitrano and Mahdi Ghelichi},
      year={2025},
      eprint={2504.06293},
      archivePrefix={arXiv},
      primaryClass={q-fin.RM},
      url={https://arxiv.org/abs/2504.06293}, 
}

@misc{distilbert,
      title={DistilBERT, a distilled version of BERT: smaller, faster, cheaper and lighter}, 
      author={Victor Sanh and Lysandre Debut and Julien Chaumond and Thomas Wolf},
      year={2020},
      eprint={1910.01108},
      archivePrefix={arXiv},
      primaryClass={cs.CL},
      url={https://arxiv.org/abs/1910.01108}, 
}

@inproceedings{WDR,
 author = {Mikolov, Tomas and Sutskever, Ilya and Chen, Kai and Corrado, Greg S and Dean, Jeff},
 booktitle = {Advances in Neural Information Processing Systems},
 editor = {C.J. Burges and L. Bottou and M. Welling and Z. Ghahramani and K.Q. Weinberger},
 pages = {},
 publisher = {Curran Associates, Inc.},
 title = {Distributed Representations of Words and Phrases and their Compositionality},
 url = {https://proceedings.neurips.cc/paper_files/paper/2013/file/9aa42b31882ec039965f3c4923ce901b-Paper.pdf},
 volume = {26},
 year = {2013}
}

@misc{WDS,
      title={Deep contextualized word representations}, 
      author={Matthew E. Peters and Mark Neumann and Mohit Iyyer and Matt Gardner and Christopher Clark and Kenton Lee and Luke Zettlemoyer},
      year={2018},
      eprint={1802.05365},
      archivePrefix={arXiv},
      primaryClass={cs.CL},
      url={https://arxiv.org/abs/1802.05365}, 
}

@inproceedings{transformer,
 author = {Vaswani, Ashish and Shazeer, Noam and Parmar, Niki and Uszkoreit, Jakob and Jones, Llion and Gomez, Aidan N and Kaiser, \L ukasz and Polosukhin, Illia},
 booktitle = {Advances in Neural Information Processing Systems},
 editor = {I. Guyon and U. Von Luxburg and S. Bengio and H. Wallach and R. Fergus and S. Vishwanathan and R. Garnett},
 pages = {},
 publisher = {Curran Associates, Inc.},
 title = {Attention is All you Need},
 url = {https://proceedings.neurips.cc/paper_files/paper/2017/file/3f5ee243547dee91fbd053c1c4a845aa-Paper.pdf},
 volume = {30},
 year = {2017}
}

@inproceedings{sbert,
  title={Sentence-BERT: Sentence Embeddings using Siamese BERT-Networks},
  author={Nils Reimers and Iryna Gurevych},
  booktitle={Conference on Empirical Methods in Natural Language Processing},
  year={2019},
  url={https://api.semanticscholar.org/CorpusID:201646309}
}

@inproceedings{bert,
    title = "{BERT}: Pre-training of Deep Bidirectional Transformers for Language Understanding",
    author = "Devlin, Jacob  and
      Chang, Ming-Wei  and
      Lee, Kenton  and
      Toutanova, Kristina",
    editor = "Burstein, Jill  and
      Doran, Christy  and
      Solorio, Thamar",
    booktitle = "Proceedings of the 2019 Conference of the North {A}merican Chapter of the Association for Computational Linguistics: Human Language Technologies, Volume 1 (Long and Short Papers)",
    month = jun,
    year = "2019",
    address = "Minneapolis, Minnesota",
    publisher = "Association for Computational Linguistics",
    url = "https://aclanthology.org/N19-1423/",
    doi = "10.18653/v1/N19-1423",
    pages = "4171--4186"
}

\appendix
\renewcommand\thefigure{\thesection.\arabic{figure}}
\setcounter{figure}{0}
\section{Appendix: Additional Similarity Search Results} \label{apndx:search}
This appendix provides additional qualitative evidence for the similarity search framework by showing how learned embeddings structure the bond universe across issuers, sectors, and regions. The examples illustrate how similarity relationships emerge, evolve with weaker issuer- and sector-level alignment, and balance fine-grained specificity with broader economic structure, demonstrating robust neighbor selection and curve reconstruction across diverse credit profiles.

\begin{figure}[]
    \centering
    \includegraphics[width=0.75\paperheight,height=\paperwidth,keepaspectratio,angle=90]{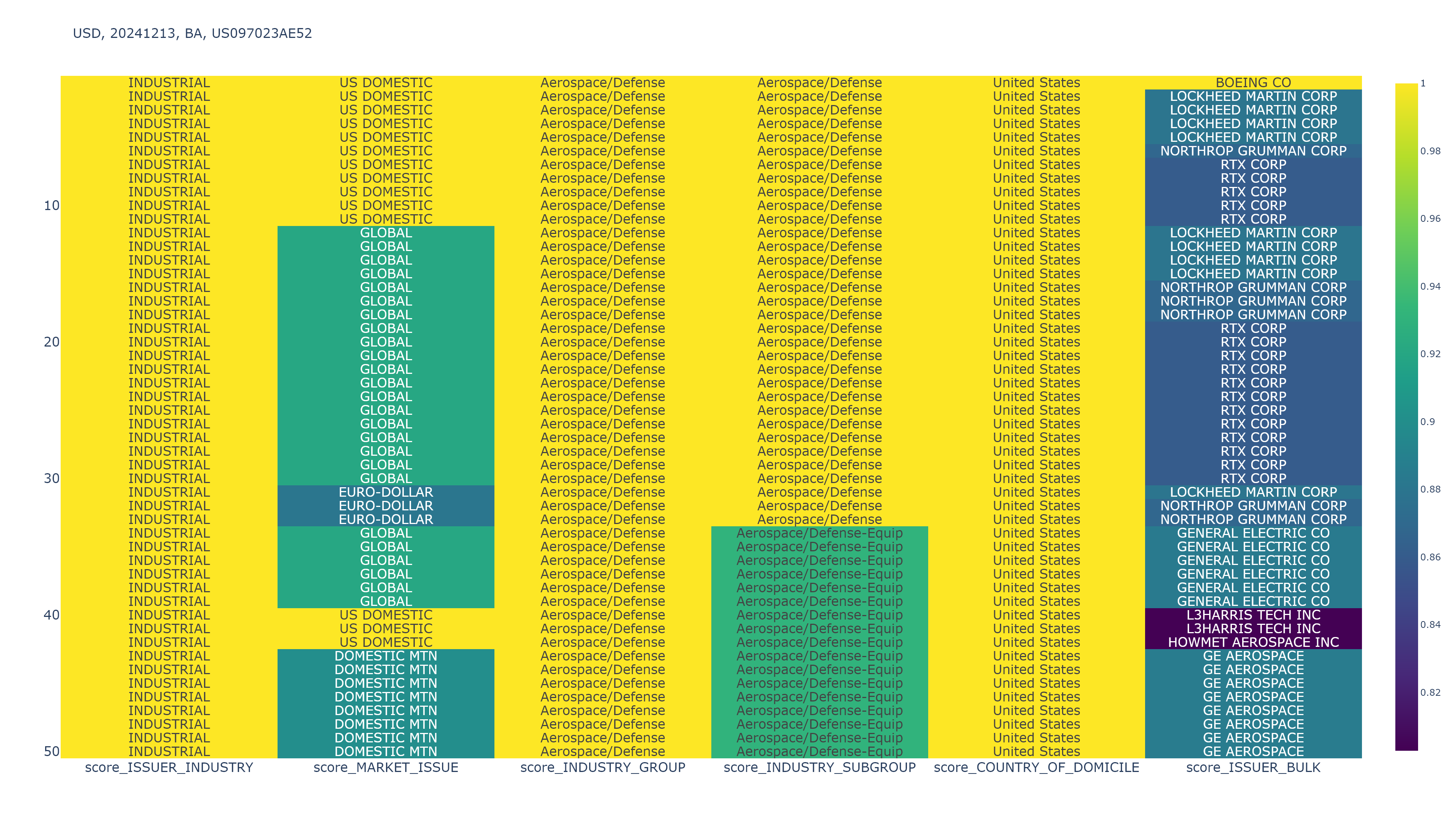}
    \caption{Visualization of bond similarity search results for the query bond of Boeing Co.}
    \label{fig:ba1}
\end{figure}

Figure~\ref{fig:ba1} presents an aerospace and defense case study, showing how the model retrieves economically coherent peers for Boeing Co. and how similarity scores decay as sector and issuer alignment weaken. It reports the top-ranked neighbors for a Boeing Co. bond, illustrating how the embedding model prioritizes issuers in the same aerospace and defense sector and domicile while smoothly degrading similarity scores as sectoral or issuer alignment weakens. The progression from Boeing to Lockheed Martin, Northrop Grumman, RTX, and related aerospace equipment names demonstrates that the learned representations preserve both issuer-level and industry-level structure, rather than relying on exact categorical matches alone.

\begin{figure}[]
    \centering
    \includegraphics[width=0.75\paperheight,height=\paperwidth,keepaspectratio,angle=90]{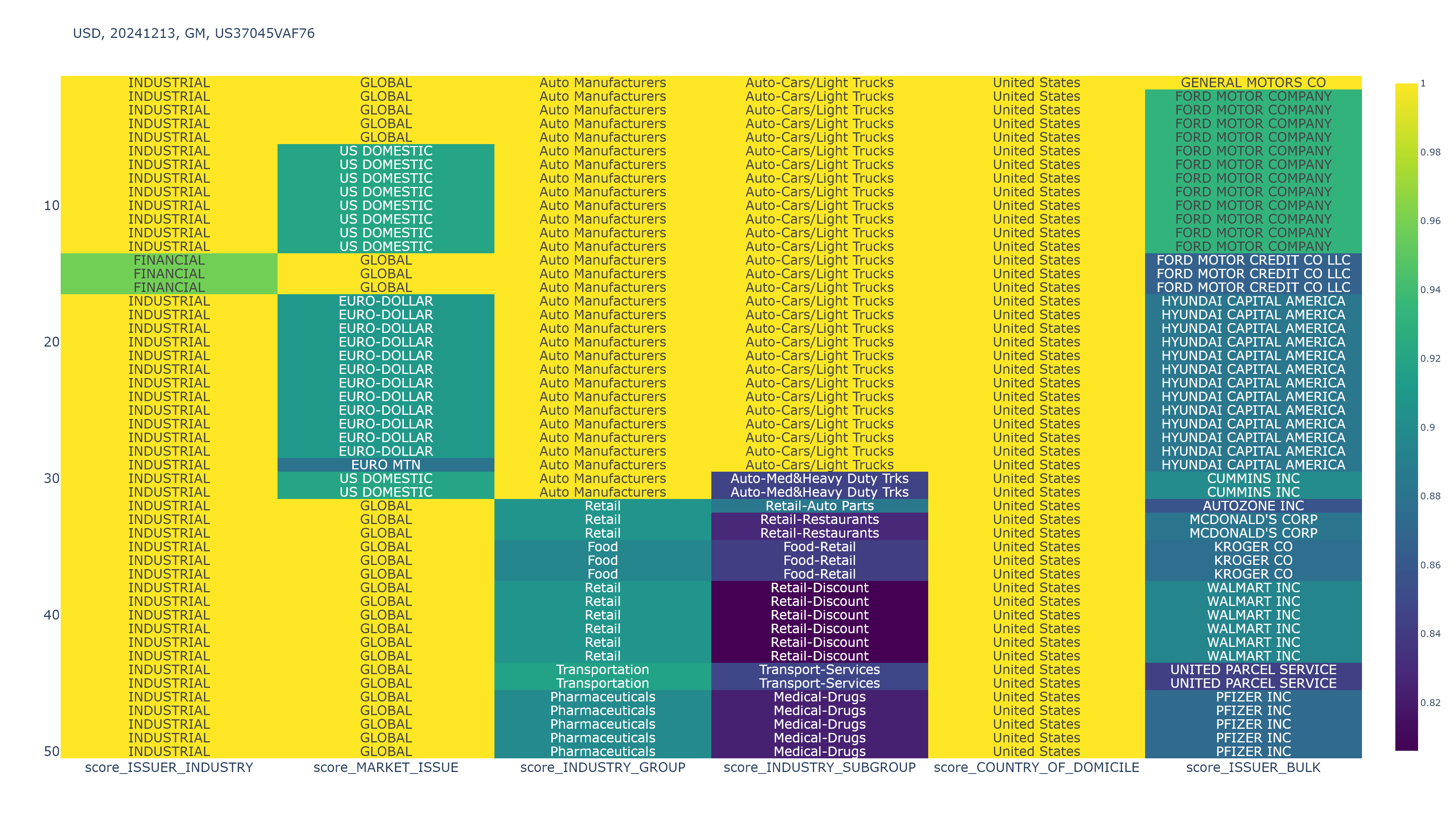}
    \caption{Visualization of bond similarity search results for the query bond of General Motors.}
    \label{fig:gm1}
\end{figure}

Figure~\ref{fig:gm1} examines an automotive and consumer-linked case, demonstrating how the model ranks General Motors neighbors along an intuitive auto retail transportation continuum. It shows the nearest neighbors for a General Motors bond, where the model first selects U.S. auto manufacturers such as Ford and related financing entities before gradually expanding to adjacent retail and transportation names. This behavior highlights the model's ability to capture second-order similarity within a broader automotive and consumer ecosystem, yielding a graded similarity profile instead of a hard boundary between ``in-sector'' and ``out-of-sector'' bonds.

\begin{figure}[]
    \centering
    \includegraphics[width=0.75\paperheight,height=\paperwidth,keepaspectratio,angle=90]{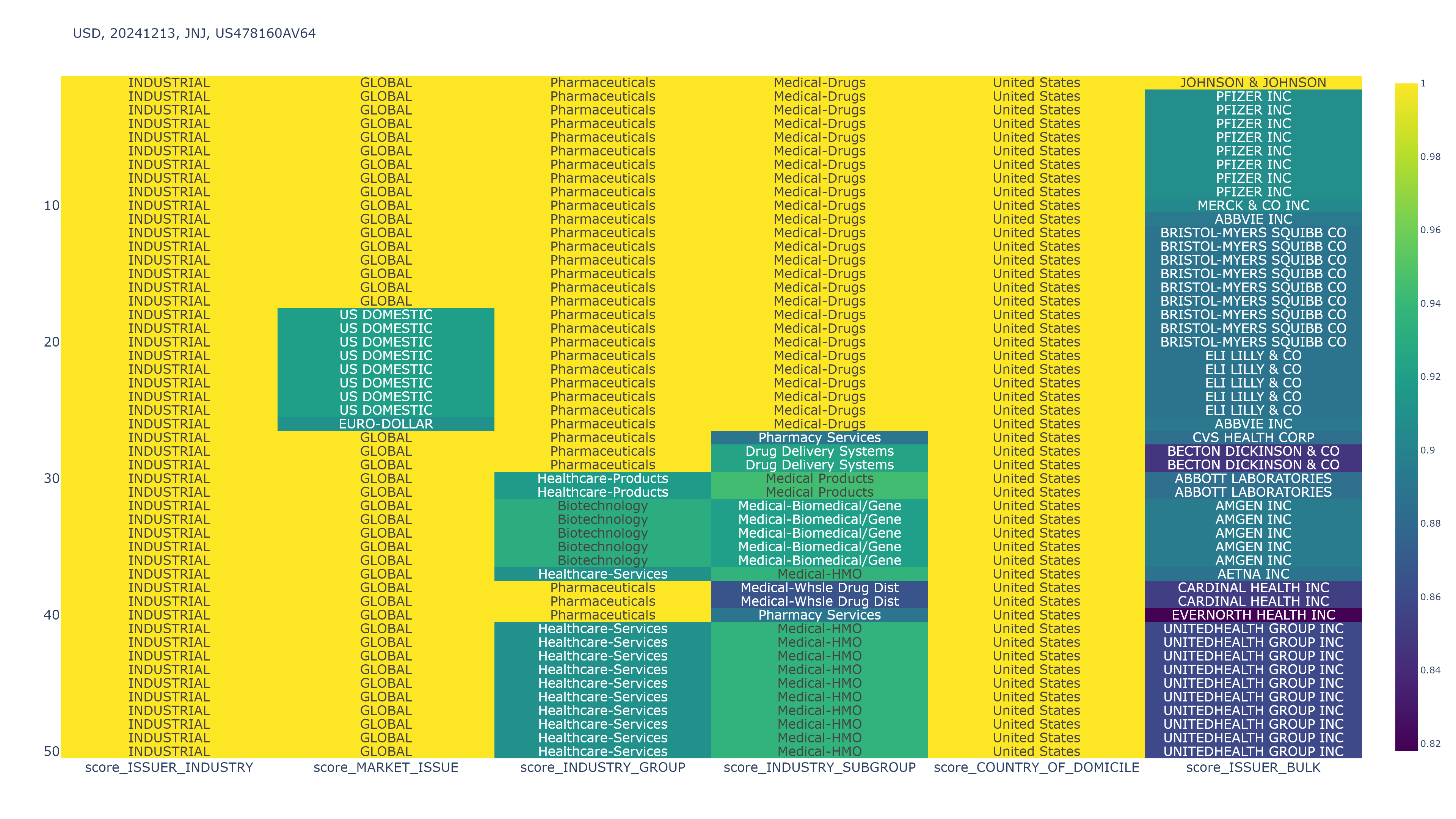}
    \caption{Visualization of bond similarity search results for the query bond of Johnson\&Johnson.}
    \label{fig:jnj1}
\end{figure}

Figure~\ref{fig:jnj1} focuses on a large-cap healthcare issuer, highlighting how the embeddings capture layered similarity within pharmaceuticals, biotechnology, and broader healthcare services. The query bond from Johnson\&Johnson is matched primarily to large U.S. pharmaceutical and healthcare issuers, including Pfizer, Merck, Bristol Myers Squibb, and Eli Lilly. The ranked list illustrates that the embedding space organizes issuers along therapeutics and healthcare-service dimensions, with similarity scores declining smoothly as the model transitions from core pharma peers to broader healthcare and insurance names.

\begin{figure}[]
    \centering
    \includegraphics[width=0.75\paperheight,height=\paperwidth,keepaspectratio,angle=90]{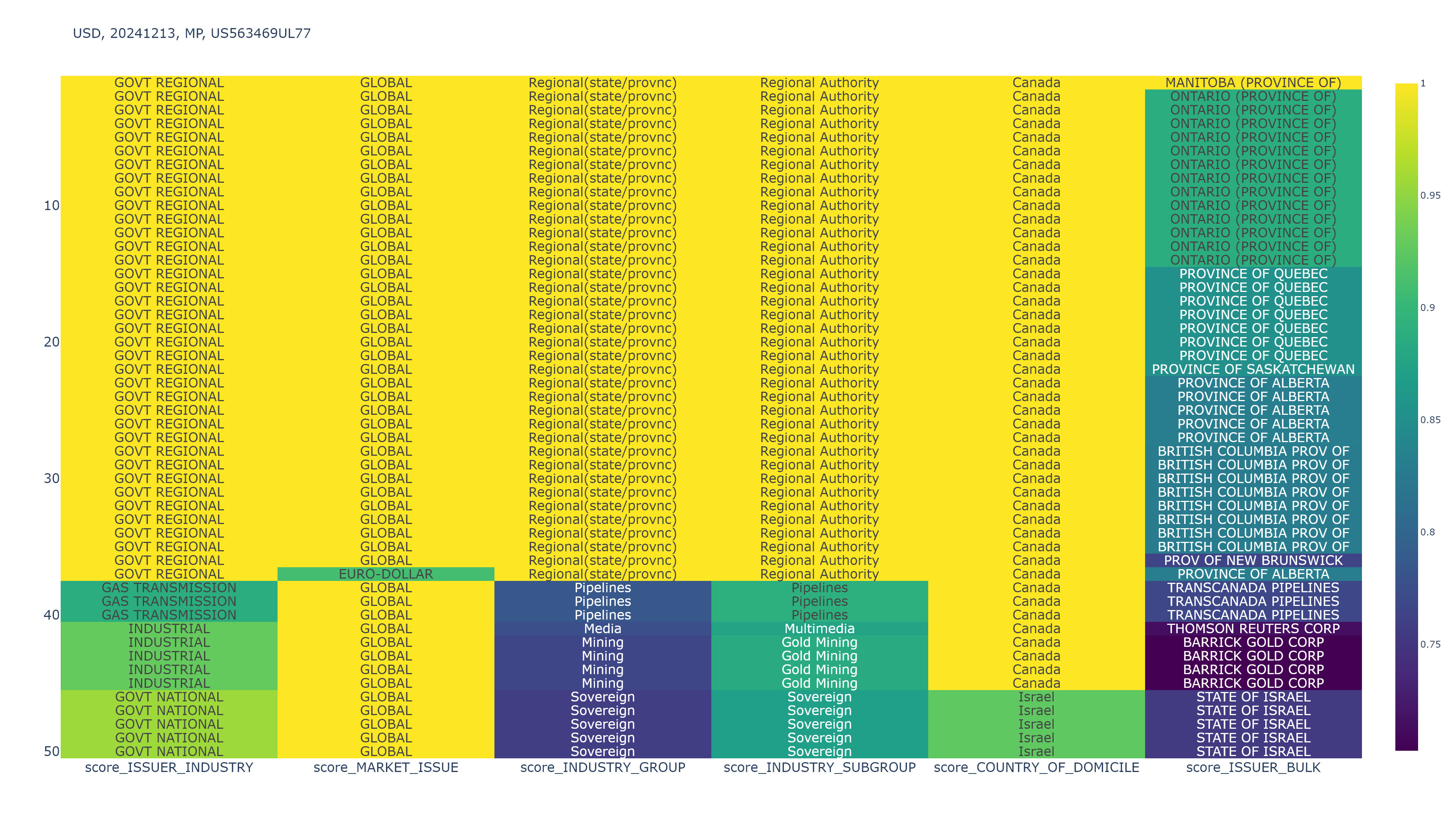}
    \caption{Visualization of bond similarity search results for the query bond of Province of Manitoba.}
    \label{fig:mp1}
\end{figure}

Figure~\ref{fig:mp1} analyzes a sub-sovereign government issuer, illustrating how the model prioritizes Canadian provincial peers before gradually expanding to pipelines, corporates, and selected sovereigns. It presents similarity search results for a Province of Manitoba bond, where the top neighbors are Canadian provincial issuers such as Ontario, Quebec, Alberta, Saskatchewan, and British Columbia. Only at lower similarity scores do non-provincial Canadian entities and selected foreign sovereigns appear, showing that the embeddings strongly prioritize regional, sectoral, and domicile alignment for sub-sovereign government bonds.

\begin{figure}[!]
    \centering
    \includegraphics[width=0.49\linewidth]{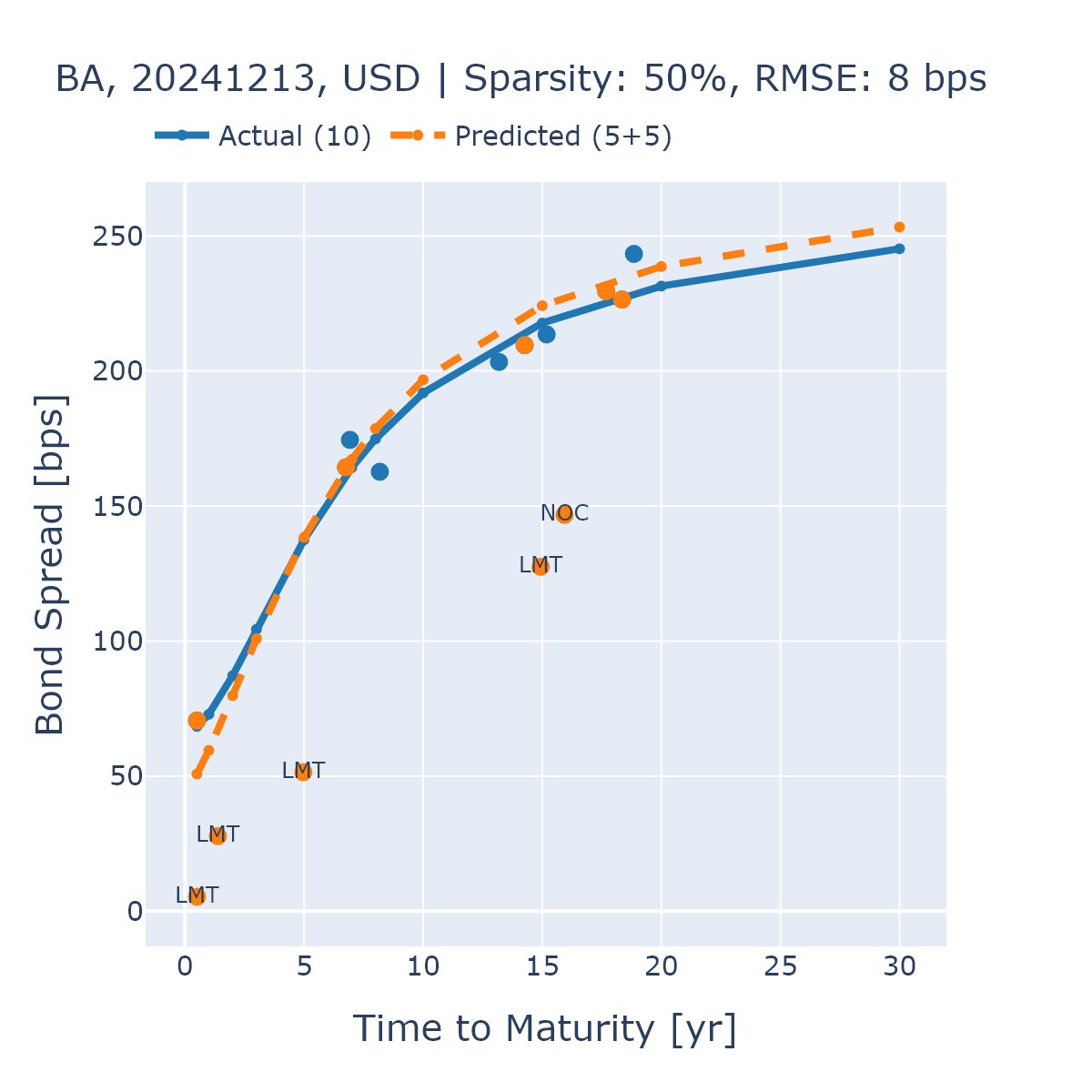}
    \includegraphics[width=0.49\linewidth]{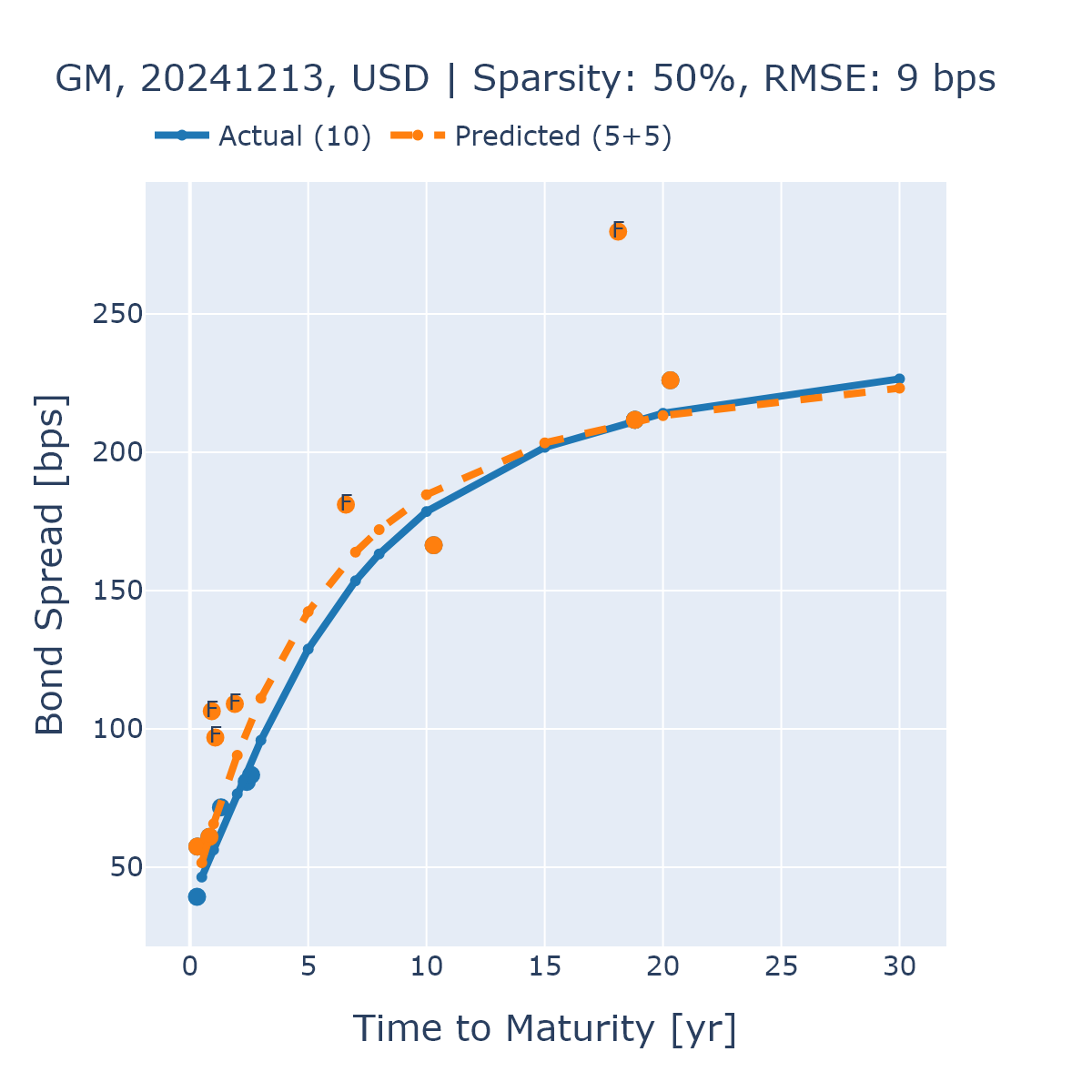}
    \includegraphics[width=0.49\linewidth]{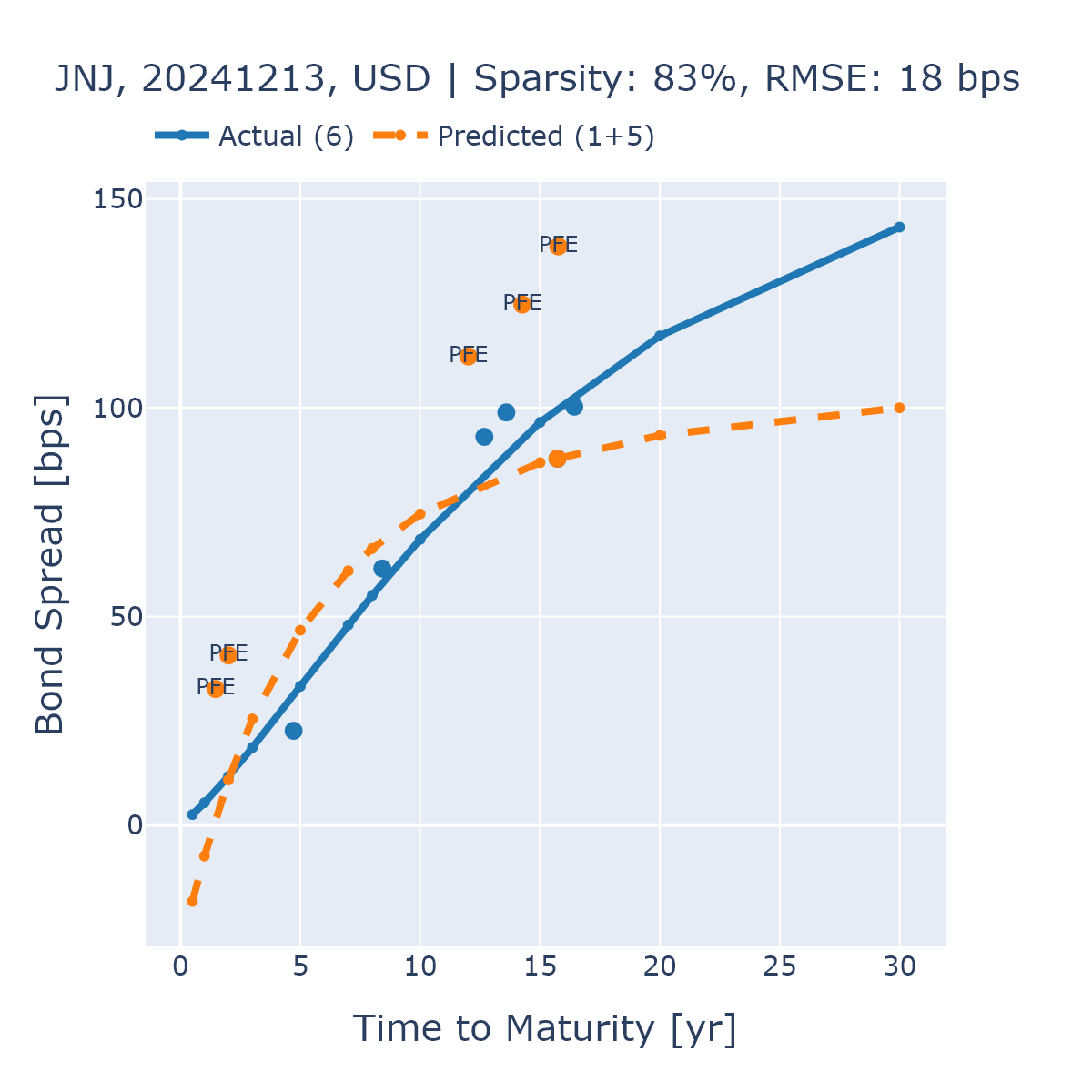}
    \includegraphics[width=0.49\linewidth]{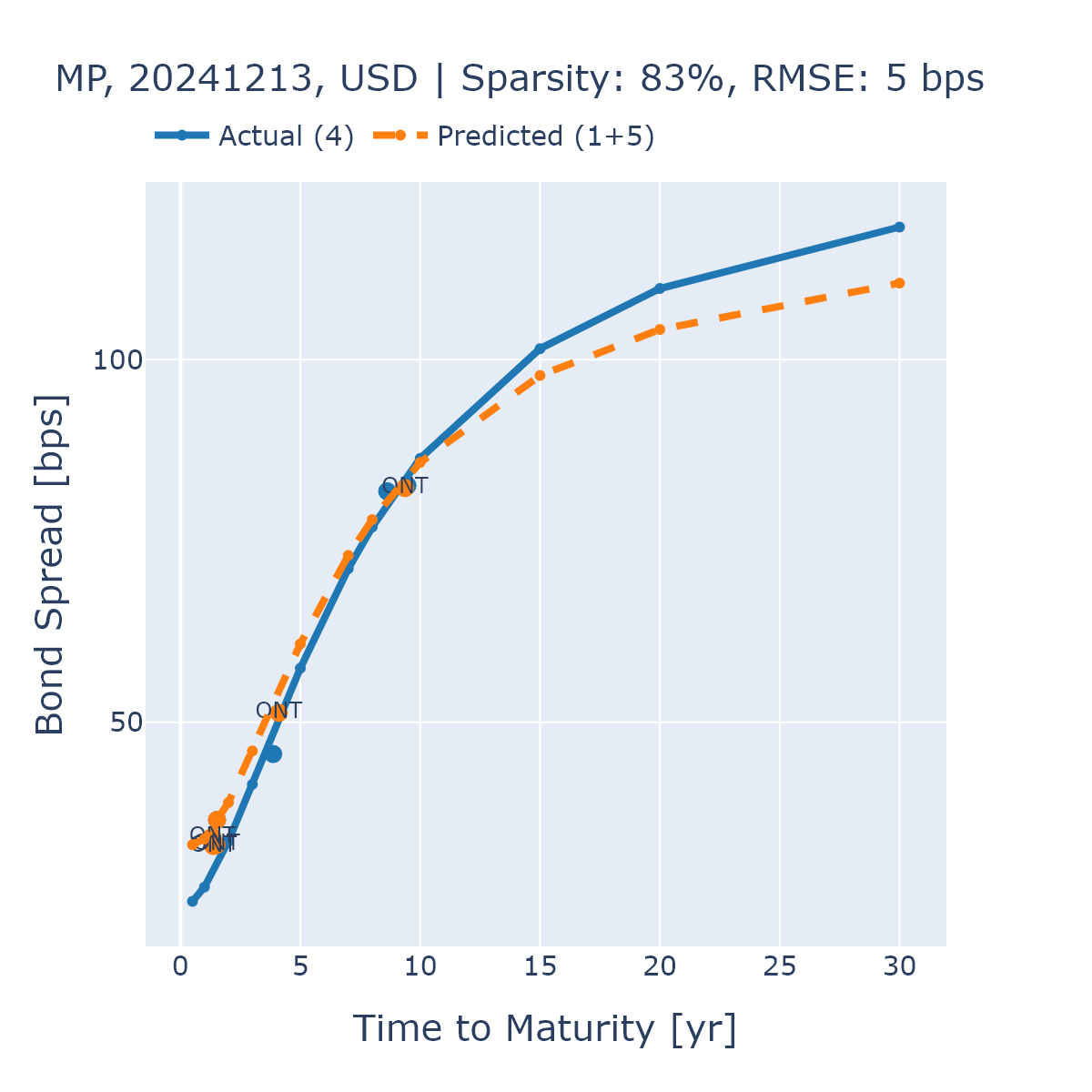}
    \caption{Comparison of predicted and actual CDS spread curves for Boeing Co. (top-left), General Motors (top-right), Johnson\&Johnson (bottom-left), and Province of Manitoba (bottom-right) issuers.}
    \label{fig:apndx0}
\end{figure}

Figure~\ref{fig:apndx0} reports CDS spread curve reconstructions for the aforementioned industrial, consumer, healthcare, and government issuers, confirming the stability of the augmentation framework across heterogeneous credit profiles. It compares actual and predicted CDS spread curves for Boeing, General Motors, Johnson\&Johnson, and the Province of Manitoba under varying sparsity levels, illustrating how reconstruction quality evolves as the available market information becomes increasingly limited. Across all four issuers, the augmented curves closely track the overall level, slope, and curvature of the observed term structures, yielding consistently low RMSEs in the range of 5--18 bps. These results demonstrate that the proposed similarity-based augmentation remains stable and effective across different sectors, even under pronounced sparsity, and that the learned representations support accurate curve reconstruction in diverse market settings.

\clearpage
\clearpage
\clearpage
\clearpage
\clearpage
\clearpage
\clearpage

\section{Appendix: Additional Benchmarking Results} \label{apndx:bench}
This appendix reports extended benchmarking experiments that compare the XEmbedding model against alternative encodings and two-step procedures, with a focus on sparsity robustness and issuer-level behavior.

\begin{figure}[!b]
    \centering
    \includegraphics[width=0.49\linewidth]{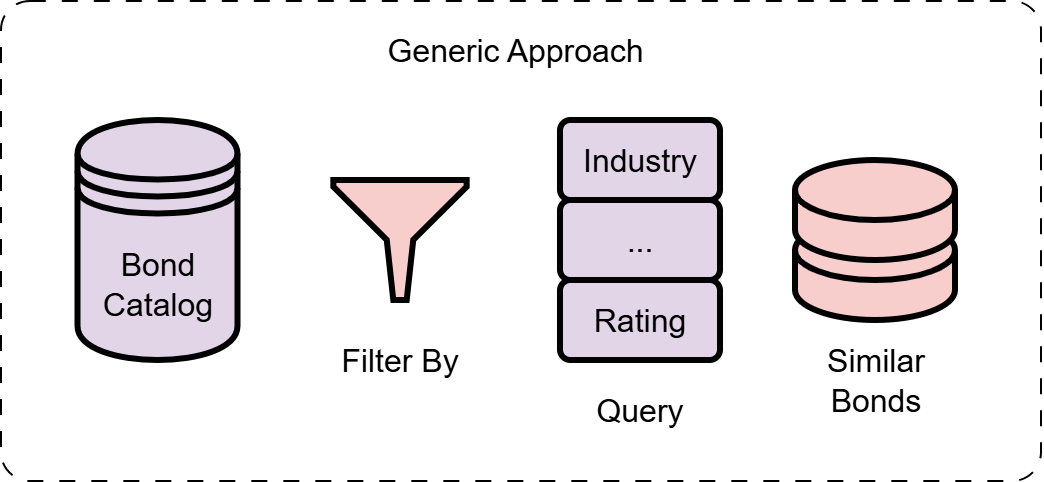}
    \includegraphics[width=0.49\linewidth]{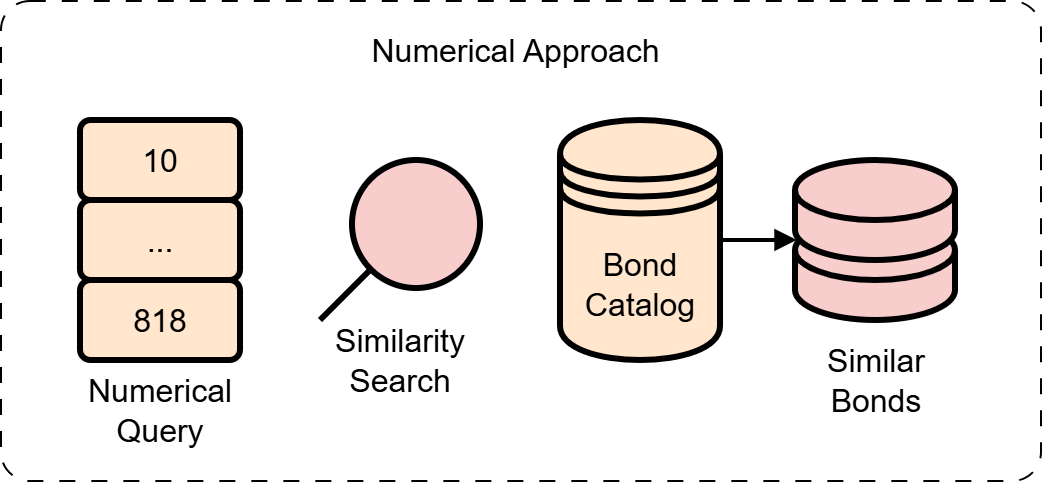}
    \includegraphics[width=0.99\linewidth]{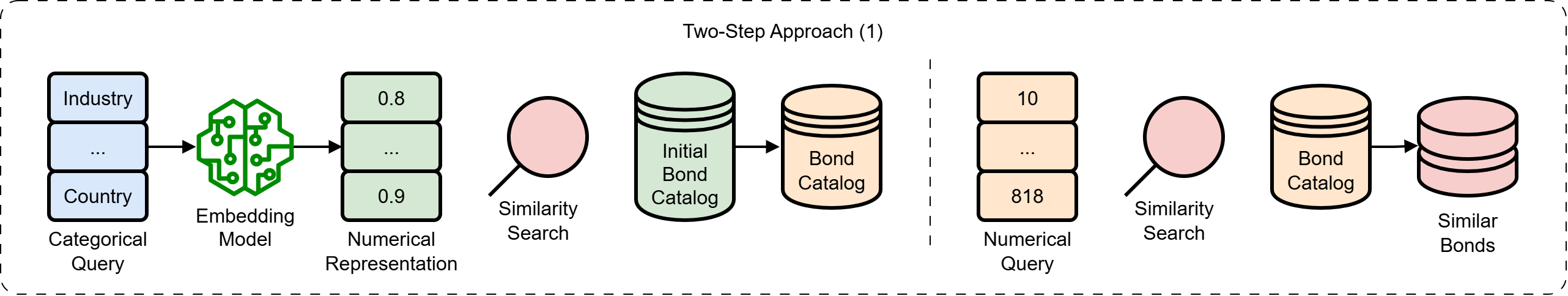}
    \includegraphics[width=0.99\linewidth]{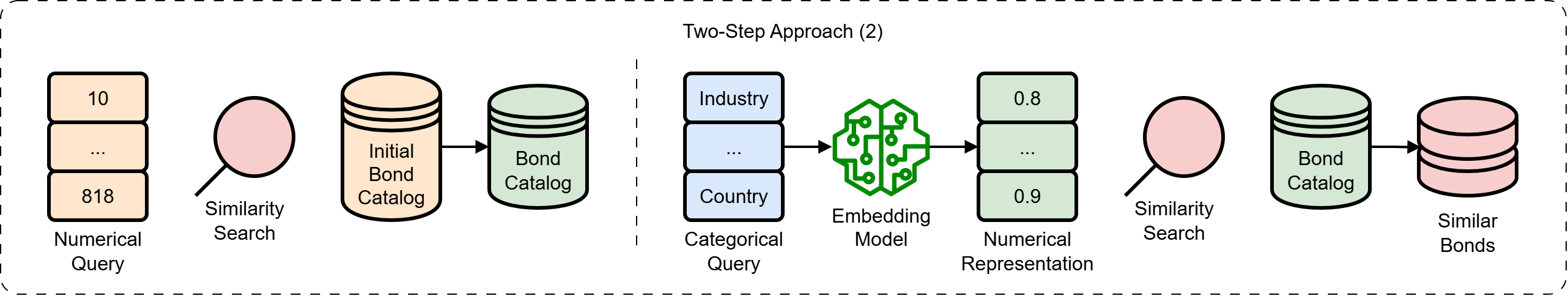}
    \caption{Comparison of alternative approaches. The generic approach (top-left) filters the bond catalog using categorical attributes such as industry and rating. The numerical approach (top-right) directly performs similarity search over numerical bond representations. The two-step approaches (bottom) combine categorical and numerical information by first narrowing the search space using one modality and then applying similarity search using the other.}
    \label{fig:alternatives}
\end{figure}

Figure~\ref{fig:alternatives} illustrates four alternative strategies for retrieving similar bonds. The generic approach relies on rule-based filtering using categorical attributes (e.g., industry or rating) to identify comparable instruments. In contrast, the numerical approach represents bonds using numerical features and retrieves neighbors directly via similarity search. To leverage the strengths of both paradigms, we introduce two two-step approaches: either categorical information is first embedded and used to construct an initial candidate set that is subsequently refined through numerical similarity search, or numerical similarity is applied first to form a reduced catalog that is then refined using categorical embeddings.

\begin{figure}[!]
    \centering
    \includegraphics[width=0.49\linewidth]{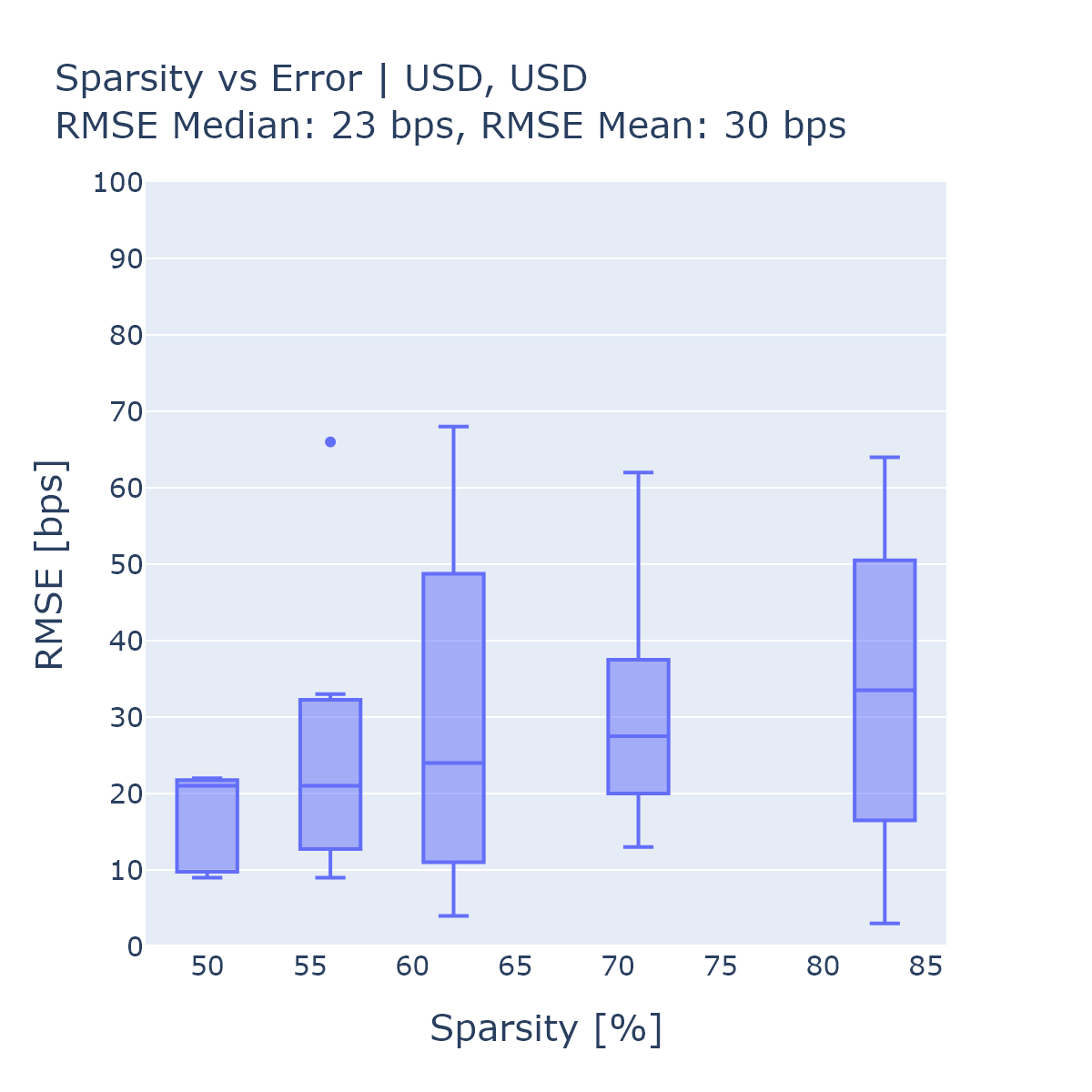}
    \includegraphics[width=0.49\linewidth]{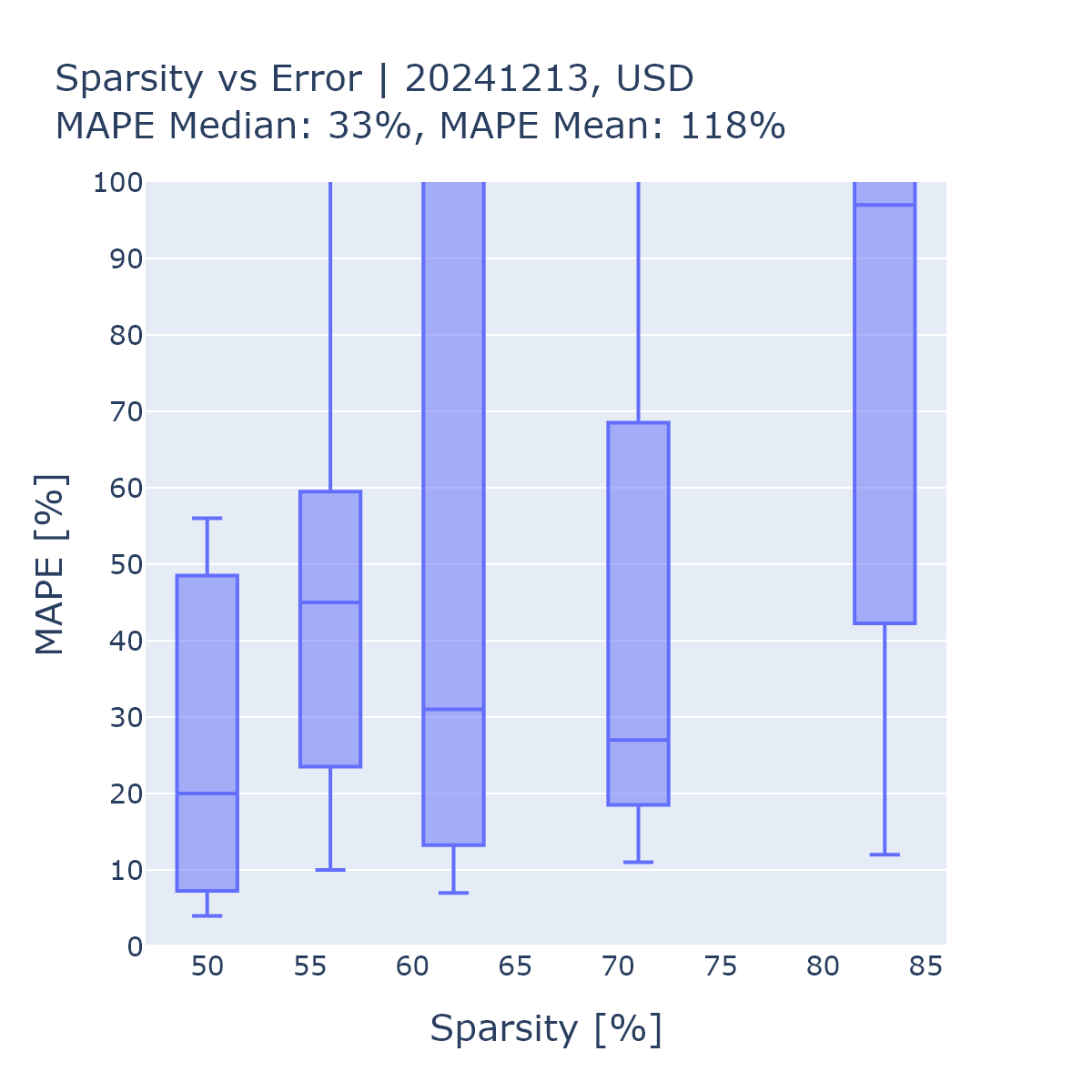}
    \includegraphics[width=0.49\linewidth]{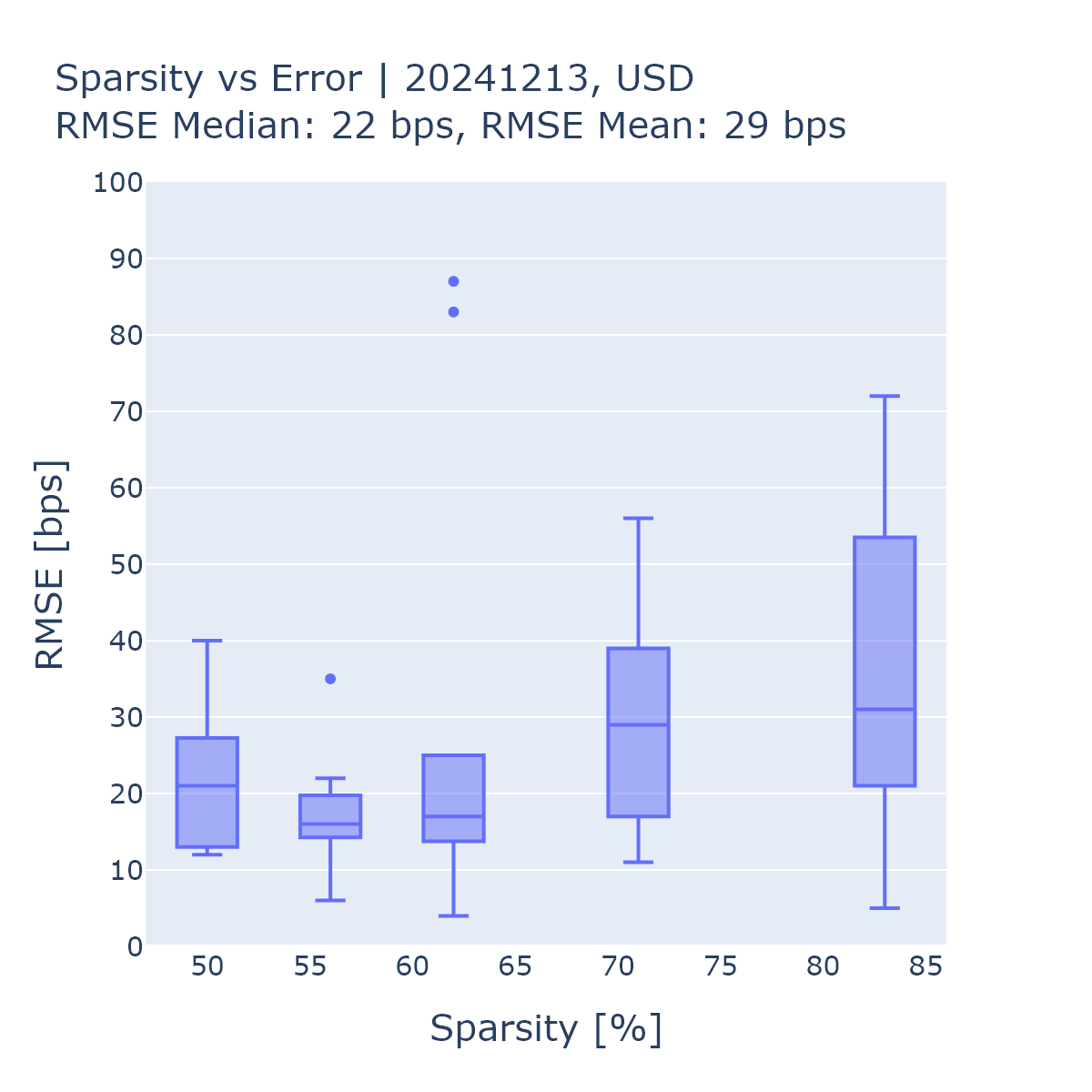}
    \includegraphics[width=0.49\linewidth]{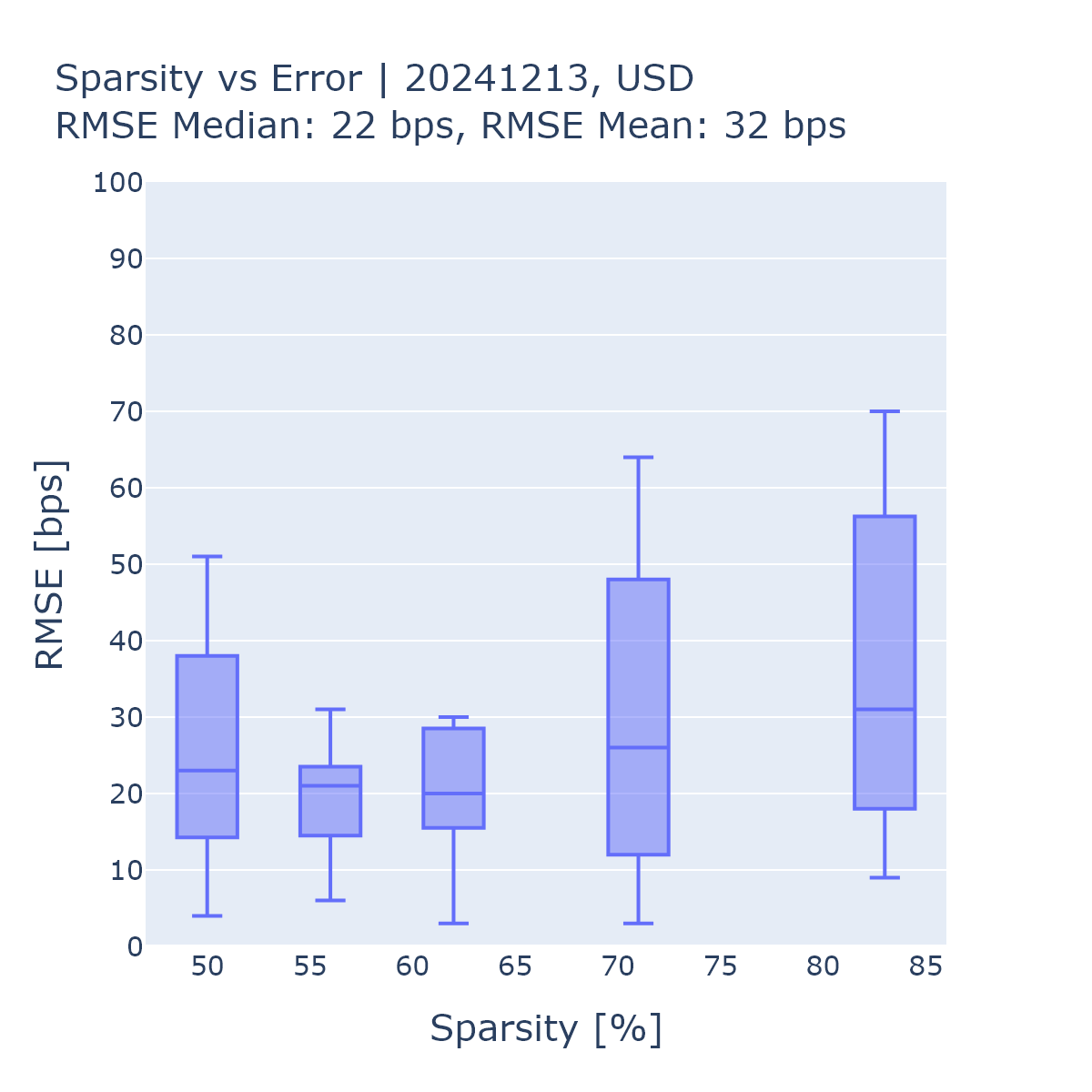}
    \caption{Comparison of overall performance of alternative approaches (top-left: Generic, top-right: Numerical, bottom-left: Two-Step (1), bottom:right: Two-Step (2)). Error as a function of sparsity for alternative baseline approaches. Boxplots summarize the distribution of prediction errors across sparsity levels, reported using RMSE (bps) and MAPE (\%). Median and mean error statistics are shown in each panel. Across all sparsity regimes, the baseline methods exhibit higher error levels and greater variability compared to the proposed XEmbedding model.}
    \label{fig:errorss}
\end{figure}

Figure~\ref{fig:errorss} reports the performance of alternative baseline approaches across different sparsity levels. As sparsity increases, both RMSE and MAPE generally deteriorate and exhibit larger dispersion, indicating reduced robustness under limited information. In contrast to these baselines, XEmbedding consistently achieves lower error across all sparsity regimes (cf. Figure~\ref{fig:errors}), with tighter error distributions and improved stability. These results suggest that explicitly learning structured representations enables XEmbedding to better exploit inputs and maintain predictive accuracy relatively.

\begin{figure}[!]
    \centering
    \includegraphics[width=0.49\linewidth]{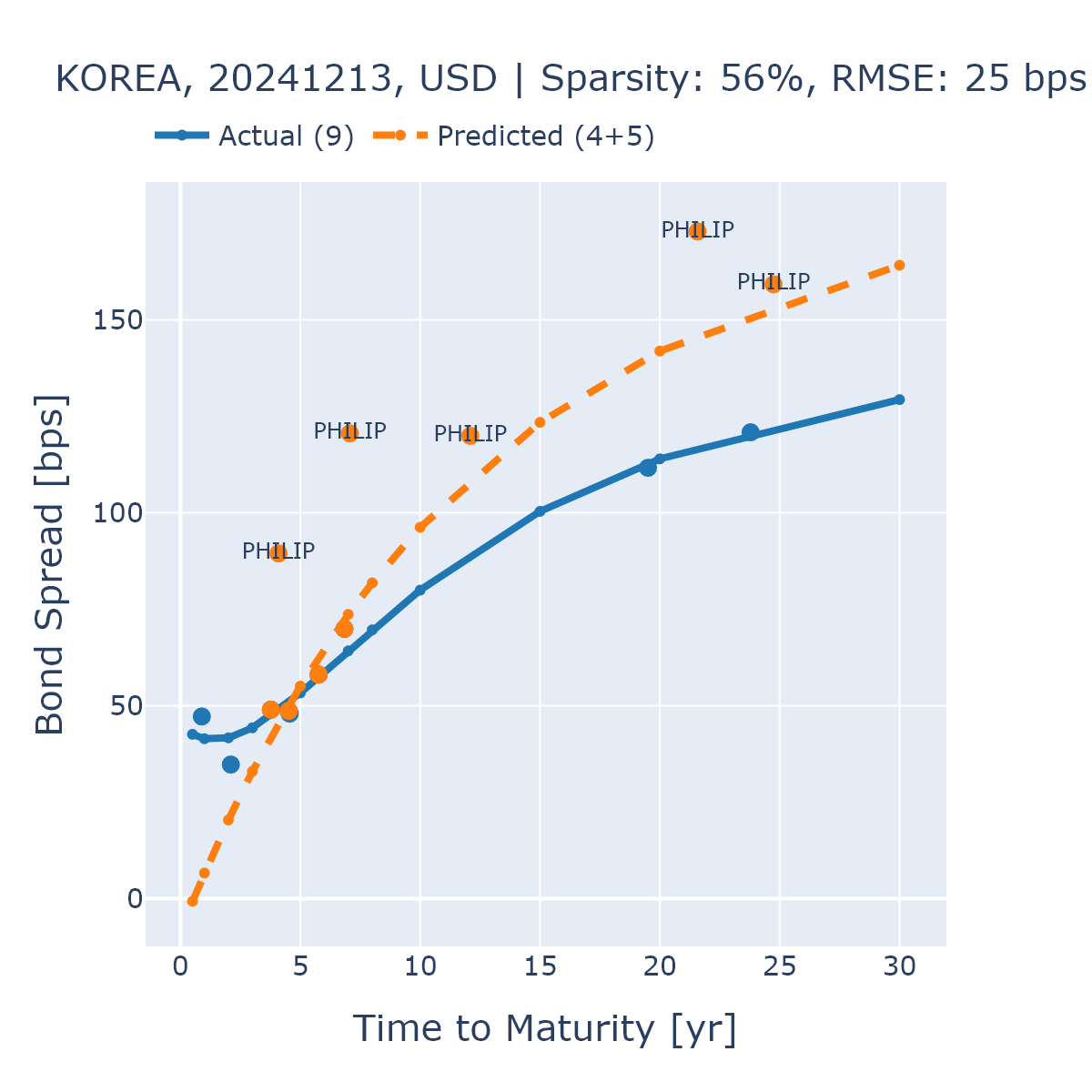}
    \includegraphics[width=0.49\linewidth]{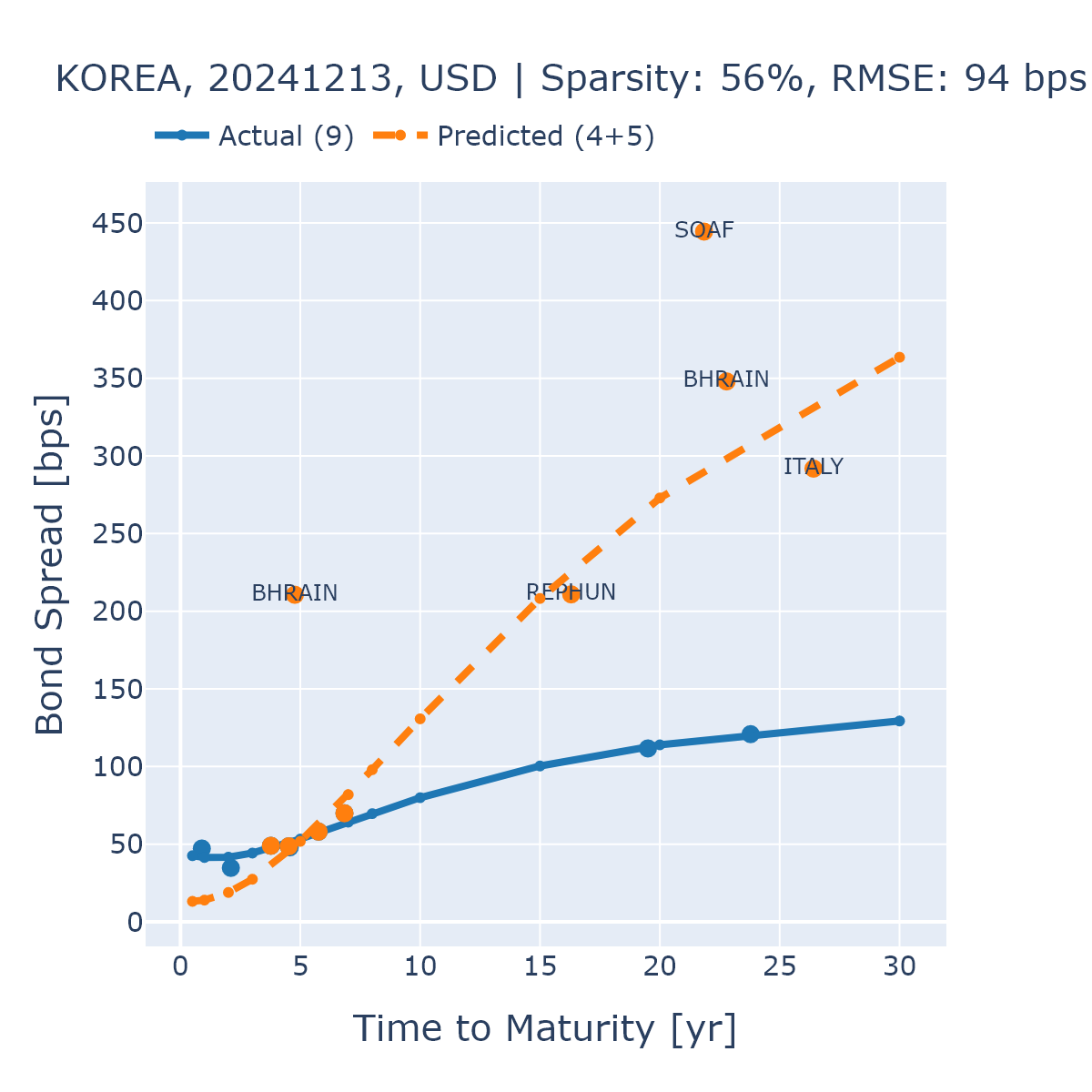}
    \caption{Comparison of XEmbedding (left) and one-hot (right) model prediction versus actual CDS spread curves for KOREA. Annotated orange markers indicate specific predicted bonds, such as PHILIP, which belongs to a similar region, and BHRAIN, which does not.}
    \label{fig:korea_onehot_xmb}
\end{figure}

\begin{figure}[!]
    \centering
    \includegraphics[width=0.49\linewidth]{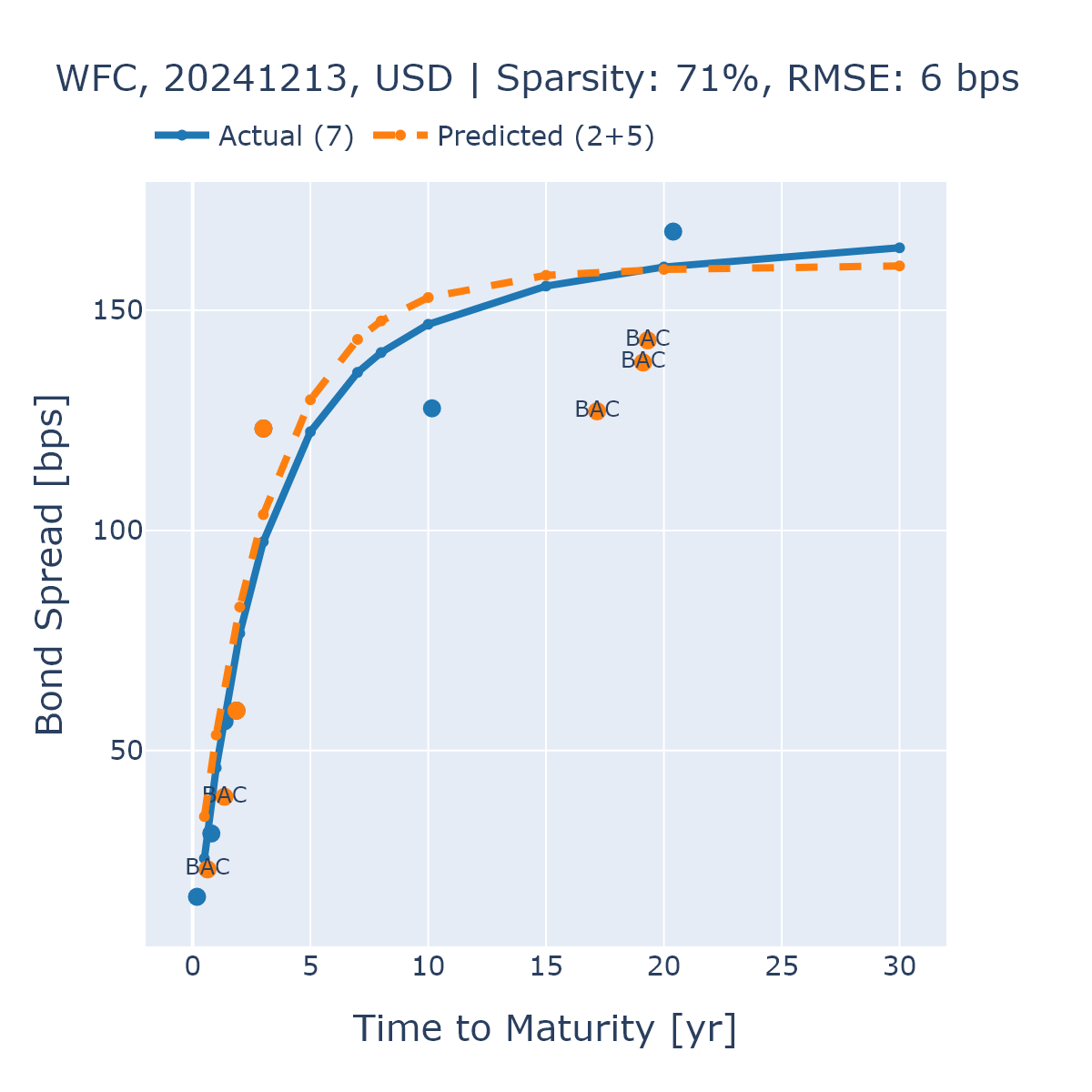}
    \includegraphics[width=0.49\linewidth]{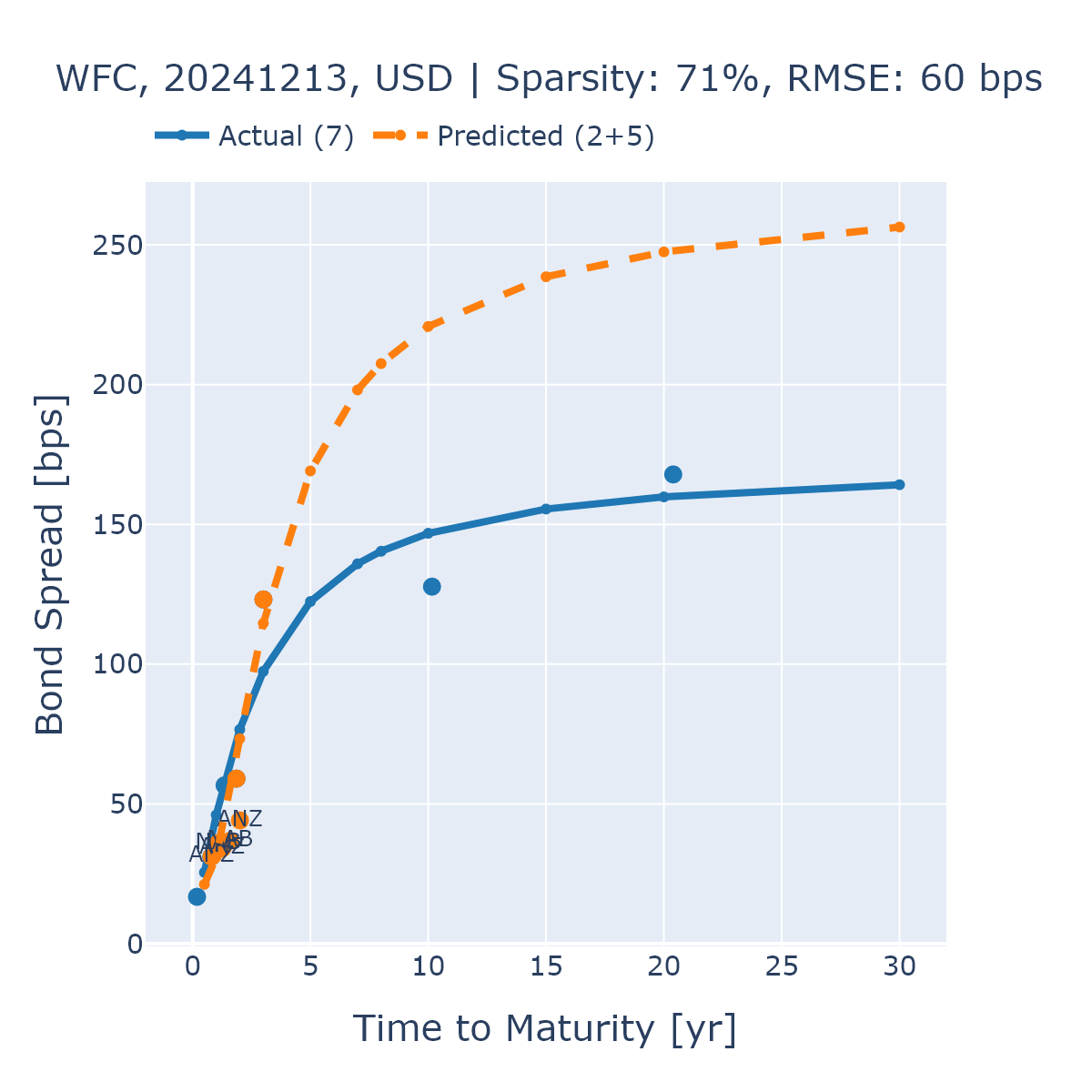}
    \caption{Comparison of XEmbedding (left) and one-hot (right) model prediction versus actual CDS spread curves for WFC (Wells Fargo \& Co). Annotated orange markers indicate specific predicted bonds, such as BAC (Bank of America Corp.), which belongs to a similar country, and NAB (National Australia Bank Ltd) and ANZ (ANZ Group Holdings Ltd), which do not.}
    \label{fig:wfc_onehot_xmb}
\end{figure}

Figures \ref{fig:korea_onehot_xmb} and \ref{fig:wfc_onehot_xmb} provide issuer-level comparisons for KOREA and WFC (Wells Fargo \& Co), illustrating how XEmbedding avoids implausible neighbors admitted by one-hot encodings and thereby achieves markedly lower RMSE and more realistic curve shapes. For both issuers, the embedding-based model selects economically plausible neighbors and delivers lower RMSEs (e.g., 6 bps vs. 60 bps), whereas the one-hot representation often admits mismatched peers from unrelated regions or sectors, leading to distorted curve shapes and inflated errors. In the case of KOREA, XEmbedding retrieves regionally aligned issuers such as PHILIP, while excluding less relevant entities like BHRAIN; similarly, for WFC, the model prioritizes country-aligned peers such as BAC (Bank of America Corp.) over geographically distant institutions including NAB (National Australia Bank Ltd) and ANZ (ANZ Group Holdings Ltd).

\clearpage
\clearpage

\clearpage
\clearpage

\end{document}